\documentclass[12pt]{article}
\input psfig.tex
\usepackage{graphicx}
\usepackage{amssymb}
\usepackage{epstopdf}
\newcommand{\ignore}[1]{}
\newcommand{\be}{\begin{equation}} \newcommand{\ee}{\end{equation}}
\newcommand{\ba}{\begin{eqnarray}} \newcommand{\ea}{\end{eqnarray}}
\newcommand{\nn}{\nonumber} \renewcommand{\bf}{\textbf}
\newcommand{\ra}{\rightarrow}

\renewcommand{\a}{\alpha} \renewcommand{\b}{\beta}

\newcommand{\p}{\partial}  
\newcommand{\EE}{\vec E}  \newcommand{\NN}{\vec \nabla}
\newcommand{\BB}{\vec B}
\begin{document}

\input{epsf}

\title{Probing Light Pseudoscalars with Light \\ 
Propagation, Resonance and Spontaneous Polarization}

\author{Sudeep Das$^a$, Pankaj Jain$^a$, John P. Ralston$^b$ and Rajib
Saha$^a$\\
$^a$Physics Department, IIT, Kanpur - 208016, India\\
$^b$Department of Physics \& Astronomy, \\
University of Kansas\\ Lawrence, KS-66045, USA\\
}

\maketitle

\noindent
\bf {Abstract:}
Radiation propagating over cosmological distances can probe light weakly
interacting pseudoscalar (or scalar) particles.
The existence of a spin-0 field
 changes the dynamical symmetries of 
 electrodynamics.
It predicts spontaneous generation of polarization 
of electromagnetic waves due to mode mixing in the presence of background
magnetic field.
We illustrate this by calculations of
propagation in a uniform medium, as well as in a slowly
 varying background medium, and finally with resonant mixing. Highly 
 complicated correlations between different
 Stokes parameters are predicted depending on the parameter regimes.  
 The polarization of propagating waves shows interesting and complex 
 dependence on frequency, the distance of propagation, coupling 
 constants, and parameters of the background medium such as the plasma 
 density and the magnetic field strength.  For the first time we study 
 the resonant mixing of electromagnetic waves with the scalar field,
 which occurs when the background plasma frequency
 becomes equal to the mass of the scalar field at some point along the
 path. Dynamical effects are found to be considerably enhanced in this 
 case. We also formulate the condition under which the adiabatic
 approximation can be used consistently, and find caveats about 
 comparing different frequency regimes.

\section{Introduction}

Light, weakly interacting spin-0 particles are predicted by many 
extensions of the Standard Model of Particle Physics. 
Such field might arise as pseudo-Goldstone bosons of some spontaneously
broken chiral symmetry (PQ) \cite{PQ,Weinberg,Dine}, 
in supergravity theories \cite{Kar,HariDass}
and the low energy string action \cite{Sen,Das,Hannestad} 
as the Kalb-Ramond field. 
The mixing of light with pseudoscalar fields in propagation has a long
history.  Karl and Clark \cite{Karl} looked for ``cosmological
birefringence'' more than 20 years ago. Pseudoscalar-electromagnetic 
field mixing has long been explored by several authors
 \cite{PS83,Maiani,HS92,RS88,Bradley,CG94,Yoshimura}.
Our focus is to explore use of polarization observables
that can separate models. Our calculations find that polarization evolution can  sometimes be highly complicated and not easily estimated by simple  
dimensional arguments.  Placing limits based on rough arguments and  
dimensional analysis is far from reliable.  Recently
Csaki et al \cite{Csaki} raised the possibility that supernova dimming  
might be
caused by photon-pseudoscalar ($\gamma-\phi$) mixing, re-igniting
interest \cite{supernovas}
in this problem. Whether or not supernovas dim, the overall
physics of propagation turns out to be quite interesting, and
the signals of polarization particularly sensitive to careful
treatment.  The variety of effects is huge.   As a result, one must  
question the validity of placing limits based simply on dimensional  
analysis or naive generalities.  Instead, the logic has to be reversed  
to seek particular effects contradicting conventional physics wherever  
they occur, and then investigate the causes.

There is another reason to consider polarization as a cosmological  
observable.  Physical theories are not tested by fitting undetermined  
parameters: they are tested by testing their symmetries.  The framework  
of  general relativity (GR) has certain symmetries by construction.   
Among these symmetries is {\it duality}, which predicts that the  
polarization of light in free space does not change.  By
the principle of Equivalence, GR then prescribes parallel transport
of polarizations that are trivial in freely falling coordinates.  It  
follows that observations inconsistent with duality might rule out  
perturbations of the metric as a false signal.   Another symmetry of  
{\it unitarity} prevents the spontaneous appearance
of polarization in an unpolarized beam of light.  If such features
contradicting GR are observed, then it rules out GR as an explanation  
of them, lifting a ``degeneracy'' of interpretation, and putting  
responsibility uniquely on something new.

There is one spectacular signal.  A spin-0   
field generically induces  polarization even in a completely  
unpolarized beam.   Indeed such a field can spontaneously polarize the  
CMB itself.  The effect
is simple and quite distinctive.  
We are emboldened to suggest that the  
{\it absence} of polarization effects  in propagation over  
cosmological distances, or in CMB derived quantities, would put stringent
limits on the existence of a light spin-0 field. In this  
regard we have eagerly awaited the release of WMAP data on CMB  
polarization while preparing this article.   We feel that the time for  
discovery with polarization observables may be ripe with this and other  
sources of data coming in the future.

\section{Symmetries of Light in the presence of spin-0 field}

The {\it existence} of a light scalar (or pseudoscalar) field 
changes the symmetries of the propagation of light.
We consider a spin-0 field $\phi$ coupled to the electromagnetic
field strength $F_{\mu \nu}$ by the action \ba S  = \int d^{4}x
\sqrt{-g} \, \Bigg[ - \frac{1}{4} F_{\mu \nu}F^{\mu \nu} + g_{\phi}  \phi \,  
\epsilon_{\mu \nu \a \b}F^{\mu \nu}F^{\a \b} \nonumber \\ 
+ g'_\phi \phi F^{\mu \nu}F_{\mu \nu}
+j_{\mu}A^{\mu} + \frac{1}{2}\p_{\mu}
\phi \p_{\mu}\phi -{1 \over 2}m_{\phi}^{2}\phi^{2}
+V(\phi)\Bigg]. \label{Ldefined}  \ea 
We include a coupling to a current  
$j_{\mu}$ for completeness, while in practice this term will lead to  
modified propagation such as plasma frequencies, etc.  For the purposes  
to be developed the potential $V(\phi)$
can be ignored as a small perturbation, and the metric $g$ be replaced
by a given background form.  
Note that we have included coupling of $\phi$ to both
$\epsilon_{\mu \nu \a \b}F^{\mu \nu}F^{\a \b}$ and 
$F^{\mu \nu}F_{\mu \nu}$, since both are allowed by the symmetries of
the theory. 
If $\phi$ is a pseudoscalar, as we assume, then the coupling $g'_\phi$
breaks parity symmetry. However this approximate
feature of the low energy world is not a symmetry of Nature, being
broken by the weak interactions physics. Even if the coupling $g'_{\phi}$
is tuned to zero at some scale, a coupling will be induced by radiative
corrections. 
In the rest of the paper, however, we ignore $g'_\phi$ since the 
limit on this
coupling is much more stringent compared to the limit on 
$g_\phi$ \cite{Carroll98}.

Free space electrodynamics has an interesting symmetry called {\it
duality}.  In the Hamiltonian density ${\cal H}$ in natural units, the
electric and magnetic fields $\vec E$ and $\vec B$ can be rotated:  \ba
   \left(\begin{array}{c} \EE' \\ \BB' \end{array}\right) &=&
\left(\begin{array}{cc}\cos \theta & \sin\theta\\ -\sin \theta & \cos
\theta \end{array}\right) \left(\begin{array}{c} \EE \\ \BB
\end{array}\right); \nn \\   {\cal H} &=& {1 \over 2} \EE^{2} + {1
\over 2} \BB^{2} \ra {\cal H} .  \ea  Although a symmetry of the
Hamiltonian density, duality is not a symmetry of the Lagrangian
density, which changes by a pure divergence.  Noether's theorem  
represents this faithfully: the duality current $K^{\mu}$ is not  
conserved, but its divergence is an identity: \ba K_{\mu}=\epsilon_{\mu  
\nu \a \b }A^{\nu}F^{\a \b}; \nn \\ \p^{\mu}K_{\mu}=\epsilon_{\mu \nu  
\a \b}F^{\mu \nu}F^{\a \b}. \ea   As a consequence of
duality symmetry, $\epsilon_{\mu \nu \a \b}F^{\mu \nu}F^{\a \b} \sim
\vec E \cdot \vec B$, which is {\it not} invariant, must be zero in a  
free-space
plane wave. Otherwise symmetry would permit rotating it into a  
different magnitude, which is a contradiction. The duality current is  
deeply related to the helicity density:  \ba \frac{1}{2}K_{0} =i \frac{  
\vec k}{\omega} \cdot \vec A_{k}^{*} \times \vec E_{k} , \ea where  
$\vec k$ is the wave number, $\omega$ the angular frequency and $\vec A  
\times \vec E $ is the canonical spin density.\footnote{Strictly  
speaking, photon spin depends on the gauge choice. The interpretation  
comes from the gauge $A^{0}=0$, which is nearly unique in having valid  
canonical spin and orbital angular momenta.}   Thus duality symmetry  
implies separate conservation of the spin and orbital angular momentum  
of light in a $\vec k$ eigenstate.   Now the coupling of $\phi$ to
$\epsilon_{\mu \nu \a \b}F^{\mu \nu}F^{\a \b}$  breaks duality  
symmetry, allowing the helicity of photons to mix, and the plane of  
polarization of light to change in propagation -- at least in the  
general case. Remarkably, the coupling of Eq. \ref{Ldefined} vanishes  
for a single plane wave: an isolated photon is ``safe''.  However light  
cannot be protected under all possible conditions, and below we explore  
the effects of a background magnetic field.

Let us contrast the $\phi$-coupling with propagation in the usual  
cosmological model.  In that event there is a coupling of light and  
gravity: the metric
enters as an ``optical medium'' in the form of \ba  S  = \int d^{4}x
\sqrt{-g} \, \left[ - \frac{1}{4} F_{\mu \nu}g^{\nu \a}(x)g^{\mu \b}(x)F_{\a
\b}\right]. \ea The energy of this system can be written \ba {\cal H} = {1
\over 2} (\, \vec D \cdot \vec E + \vec B \cdot \vec H\,) ; \nn \\
D_{k}=  g^{\nu 0}(x)g^{\mu k}(x)F_{\mu \nu}; \nn \\  H_{k}=
\frac{1}{2} \epsilon_{ijk}g^{\nu j}(x)g^{\mu k}(x)F_{\mu \nu}. \ea This
system has a new duality of rotating $\vec D $ and $\vec H$ into $\vec
E$ and $\vec B$. In a coordinate system where the metric is trivial,
then $\vec D = \vec E$ and $\vec H=\vec B$ and the state of
polarization is preserved.

GR also maintains a symmetry under which the power transmitted scales  
with
kinematic dependence on the scale factor, while the plane and  
magnitudes of
polarizations are preserved, in the sense of parallel transport along
geodesics.  The prediction of GR effects then consist of a coordinate  
transformation.  In the isotropic, homogeneous metrics assumed in  
cosmology, the observable result is a red-shift.

Free-space symmetries map into GR symmetries due to the principle of  
equivalence. Thus conclusions about them do not hinge on particular  
solutions.  With GR serving as a ``null'' theory, the effects of
$\gamma-\phi$ mixing can be separated and distinguished: simply observe  
the spontaneous polarization of light during propagation, for instance.

\subsection{Mixing Light with Pseudoscalars}

Astrophysics constraints limit the coupling $g_{\phi}$ to be very tiny,  
yet the enormous length scales of cosmology allows small cumulative  
effects to develop into large ones. Proceeding requires discussion of  
scales so that we can linearize Eq. \ref{Ldefined} to discuss mixing of  
modes in propagation.  There are two important classes: {\it background  
$\phi$}, where the electromagnetic field is solved with a fixed  
pseudoscalar field, and {\it background $\vec B$}, where a combination  
of the electromagnetic field and $\phi$ are solved for a given magnetic  
field. Here we set the background $\phi$ field to zero and  put our  
effort on the second case.

We will use symbol $\vec {\cal B}$ for the magnetic background, and  
$\vec B$ for the field in propagation: the total magnetic field is  
$\vec {\cal B}+ \vec B$.  Although galactic and intergalactic $\vec  
{\cal B} \sim \mu G $ fields are small, their effects are likely to be  
large compared to the effects of $\phi$.  A quick calculation shows  
that $\vec E/c$ and $\vec B$ of a typically weak but observable signal,  
such as light from a QSO or the CMB, are also small compared to $\mu  
G$.  We seek to linearize consistently with $\vec E/c<< \vec {\cal B},$  
and $\vec B <<\vec {\cal B}$.   There is a well-defined 3-state propagation problem of mixing 2 light  
polarizations with coupling to $\phi$. We first obtain the  
non-covariant form of Maxwell equations, as follows, and with no  
approximations:  \begin{eqnarray}
\nabla\cdot {\vec E} &=& g_{\phi}\nabla\phi\cdot{(\vec {\cal B}+ \vec B  
)} +\rho ;\label{emeq1}
\\
\nabla\times{\vec E} + {\partial {(\vec {\cal B}+ \vec B )}\over  
\partial t} &=& 0 ;\label{emeq2} \\
  \nabla\times{\vec B} - {\partial {\vec E}\over \partial
t}
&=& g_{\phi} \left({\vec E}\times \nabla \phi - {  (\vec {\cal B}+ \vec  
B )}
{\partial \phi\over \partial t}
\right)  + \vec j ; \label{emeq3}
\\
\nabla\cdot{  (\vec {\cal B}+ \vec B )}&=& 0.
\label{emeq4}
\end{eqnarray}  Here ${\cal B}_{i}+B_{i}  
=\frac{1}{2}\epsilon_{ijk}F^{jk}$ and
$E_{i}=F^{0i}$ are the usual electric and magnetic fields.  Besides
this we have the equation for the pseudoscalar field
\begin{equation}
{\partial^2 \phi\over \partial t^2} - \nabla^2\phi + m_\phi^2\phi =
-g_{\phi}{\vec
E}\cdot
(\vec {\cal B}+ \vec B )
\label{pseudoscalar}
\end{equation}  Gauge invariance is explicit and one can check current  
conservation directly, $$ \NN \cdot \vec j+ {\p \rho \over \p t}=0. $$  
The dynamics of this coupled 3-field system are complicated and been  
visited by various approximations in the
literature.

We assume $\vec {\cal B}$ solves the zeroeth order Maxwell equations  
with no $\phi$ background.  Our linearized equations for $\vec E/c<<  
\vec {\cal B}, \, \vec B <<\vec {\cal B}$ are  \begin{eqnarray}
\nabla\cdot {\vec E} &=& g_{\phi}\nabla\phi\cdot{\vec {\cal B} } +\rho;  
\label{remeq1}
\\
\nabla\times{\vec E} + {\partial \vec  B \over \partial t} &=& 0;  
\label{remeq2} \\
  \nabla\times{\vec B} - {\partial {\vec E}\over \partial
t}
&=& -g_{\phi}     \vec {\cal B}
{\partial \phi\over \partial t}
    + \vec j ; \label{remeq3}
\\
\nabla\cdot{ \vec B }&=& 0. \label{remeq4}
\end{eqnarray} These equations remain exactly consistent with current  
conservation, $ \NN \cdot \vec j+ \p \rho / \p t =0. $

Proceed to get a wave equation for $\vec E$ by taking the curl of  
Faraday's Law, 
and  
substituting into Eqs. \ref{emeq1}, \ref{emeq3}.  Replacing $\vec B \sim  
\vec {\cal B}$ gives  \ba
-\NN^{2 }\vec E + {\p^{2} {\vec E} \over \p t^{2}  } = g_{\phi} \vec  
{\cal B}{\p^{2} \phi \over \p t^{2}  }-g_{\phi}  \NN ( \NN\phi \cdot  
\vec {\cal B} ).  \label{Ewave} \ea  In this equation the longitudinal  
part of $\vec E$ mixes with $ \NN\phi \cdot \vec {\cal B}$. That is,  
$\vec E$ is not perfectly transverse in general when coupled to $\phi$.

Continuing along these lines towards a general solution greatly  
increases the complexity of the equations.  We will be content here to  
show illustrative calculations restricted to the limit $ \NN\phi \cdot  
\vec {\cal B}=0.$  In this case $\vec E$ is transverse and the system  
readily collapses to 2-state mixing.  We have extensive calculations  
for the more general cases, but we feel this paper would not be  
improved by introducing long and complex calculations.
Our detailed calculations show that the numerical importance
of $B_L \neq 0$ effects for the parameter regions considered in this 
paper are negligible. The only exception to this may possibly be
in the case of resonance where the correction terms 
may be lumped together with other
parameters in any event.
The virtue of $  
\NN\phi \cdot \vec {\cal B}=0,$ even if it is not the most general case in  
practice, is that one can get a feel for the rich interplay of several  
dimensionful scales including $g_{\phi}, \, {\cal B}, \, m_{\phi}^{2}$,  
the plasma frequency $\omega_{p}$, and the propagation distance $z$.  
The amazing range of phenomena one should seek observationally is well  
illustrated.

\paragraph{Requirement 0:  Existing Physics}

Known physics of electromagnetic propagation in matter must be taken  
into account. It is well known that the plane of linearly polarized  
light rotates in a propagation through magnetized plasma: the Faraday  
effect.  Our symmetry arguments are consistent, because duality is  
broken by the coupling to electric (as opposed to magnetic) charges and  
currents.   Fortunately Faraday rotation is quite frequency dependent,  
and most
important for low frequencies below the GHz regime.  

We also take into account the practical ``photon mass'' in the form of  
the plasma frequency.  It is not an effect that can be ignored, and we  
do incorporate it in the momentum-space propagation equations below.

\paragraph{Requirement 1: Phase Accuracy}

Observables depend on the wave number differences $\Delta k$
accumulating phase $\Delta \theta$ by propagating over large distances
$\Delta z$.  An accurate approximation needs the {\it absolute phase  
error}
(symbol $\delta$) to be small compared to $\pi$, namely $\delta ( \,
\Delta k_{\pm} \Delta z  \,) <<1$.  This requirement becomes more and
more demanding as $\Delta z \ra \infty$. Everyone recognizes
this requirement, listed here for completeness. In the absence of
degeneracies, the wave numbers of propagating waves can often be  
obtained consistently in powers of $g_{\phi}$, provided the corrections  
are controlled: see below.

\paragraph{Requirement 2: Gauge Invariance}

It is very important for all symmetries to be respected, and in
particular, any violation of gauge invariance is unacceptable.   Fortunately our equations set up in  
a gauge invariant manner coincide with the standard literature in the  
limit of $\vec B \cdot \vec {\cal B}=0$.

\paragraph{Requirement 3: Respect for Limit Interchange}  Shortly we  
will solve the propagation and present a common series expansion in  
$g_{\phi}^{2}$ to simplify the formulas.  We will find that the series  
contains factors of the frequency $\omega$.  The limit of  
$g_{\phi}^{2}$ fixed and small, and taking $\omega \ra \infty$, does  
not commute with $\omega$ fixed and $g_{\phi}^{2} \ra 0$.  Since the  
frequency of a eV photon in units of the cosmological length scale is  
huge, while the coupling constants contemplated for $\phi$ is tiny, one  
needs to examine series expansions carefully to be sure they apply to  
the problem being solved.

\section{Pseudoscalar-Photon Mixing: Uniform Background}

Choose the coordinate system with the $z$-axis along the direction of  
propagation of the wave and the $x$-axis parallel to the transverse  
component
of the background magnetic field $\vec {\cal B_T} $.  Seek solutions  
with harmonic time dependence $e^{-i \omega t}$.  Denote the component  
mixing by subscript $ \parallel$ and its perpendicular complement by  
subscript $ \perp$.  Define symbol $\vec A =\vec E/\omega$, which simplifies  
equations much like the vector potential, except there is no ``i'' and  
$\vec A$ is gauge invariant.

The perpendicular component
$ A_\perp$ does not mix with $\phi$.  The wave equation for the mixing  
of $A_{||}$ and $\phi$ can be written as,
\begin{equation}
(\omega^2 + \partial_z^2)\left(\matrix{A_\parallel(z)\cr \phi(z)}\right)
- M \left(\matrix{A_\parallel(z) \cr \phi(z)} \right)= 0
\label{eq:mixing}
\end{equation}
where the ``mass matrix'' or ``mixing matrix'' is
\begin{equation}
M = \left(\matrix{\omega_p^2 & -  g_{\phi}{\cal B_T}\omega\cr  -  
g_{\phi}{\cal B_T}\omega & m_\phi^2}\right)\ ,
\label{eq:massmatrix}
\end{equation} and ${\cal B_T}$ is the magnitude of the
vector $\vec {\cal B_T}$.   At this point we took into account the  
plasma frequency  $\omega_{p}^{2}$, which is a (non-local) gauge  
invariant mass for $\vec
E$.  Transform to a new basis
\begin{equation}
\left(\matrix{A_\parallel(z)\cr \phi(z)}\right) = O
\left(\matrix{\overline{A}_\parallel(z)\cr \overline{\phi}(z)}\right)
\end{equation}
where $O$ is the orthogonal matrix
\begin{equation}
O = \left(\matrix
{\cos\theta & -\sin\theta\cr \sin\theta & \cos \theta } \right)
\end{equation}
which diagonalizes the mixing matrix in Eq. \ref{eq:mixing}.
Diagonalization gives
\begin{equation}
\tan 2\theta = {2g_{\phi}\omega {\cal B_T}\over m_\phi^2 - \omega_p^2}
\label{eq:tan2theta}
\end{equation}  with mass eigenvalues
\begin{equation}
\mu_\pm^{2} = {\omega_p^2 +  m_\phi^2\over 2} \pm {1\over 2}
\sqrt{ (\omega_p^2-m_\phi^2)^2+ (2g_{\phi}{\cal B_T}\omega)^2 }\ .
\end{equation}  We list the leading order expansion in  
$g_{\phi}$, by which \ba   \mu_{+}^{2} =  \omega_{p}^{2 } + \frac{  
g^{2} {\cal B_{T}}^{2} \omega^{2} }{\omega_{p}^{2}  - m_{\phi}^{2}   }  
; \\    \mu_{-}^{2} = m_{\phi}^{2 } - \frac{ g^{2} {\cal B_{T}}^{2}  
\omega^{2} }{\omega_{p}^{2}  - m_{\phi}^{2}   }. \label{slop} \ea We  
noted earlier that examination of scales is needed to justify this  
step.  Consistency requires \ba  \frac{ g^{2} {\cal B_{T}}^{2}  
\omega^{2} }{ (\omega_{p}^{2}  - m_{\phi}^{2})^{2}   } <<1   
\label{smallg}. \ea

Continuing, the equations are written in the diagonal basis as
\begin{eqnarray}
(\omega^2+\partial_z^2)\overline A_\parallel -
\mu^2_+\overline A_\parallel &=& 0\nonumber\\
(\omega^2+\partial_z^2)\overline \phi - \mu^2_-\overline \phi & =& 0
\label{eq:diagonalized}
\end{eqnarray}
which can be easily solved to give
\begin{eqnarray}
\overline {A}_\parallel(z) &=& \overline{A}_\parallel(0)  
e^{i(\omega+\Delta_A)z}\nonumber\\
\overline {\phi}(z) &=& \overline{\phi}(0) e^{i(\omega+\Delta_\phi)z}
\end{eqnarray}
 and
\begin{eqnarray}
\omega+\Delta_A = \sqrt{\omega^2-\mu^2_+}
\approx \omega -
{\mu^2_+\over 2\omega}\nonumber \\
\omega+\Delta_\phi = \sqrt{\omega^2- \mu^2_- }
\approx
\omega -{\mu^2_-\over 2\omega}
\label{eq:phases}
\end{eqnarray}
We will always work in the limit $\omega^2>>\mu^2_\pm$,
justifying retention of just the leading
terms. The phase difference $\Delta_\phi - \Delta_A$ in this limit
is found to be
\begin{equation}
\Delta_\phi - \Delta_A \approx {1\over 2\omega} \sqrt{(\omega_p^2 -  
m_\phi^2)^2
+(2g_{\phi}{\cal B_T}\omega)^2}
\end{equation}
The perpendicular component $A_\perp$ does not mix with $\phi$
and is given by
\begin{equation}
{A_\perp}(z) = {A_\perp}(0) e^{ik_{0}z}
\end{equation}
where $k_{ 0}= \sqrt{\omega^2 - \omega_p^2} \sim \omega -  
\omega_{p}^{2}/2 \omega $.

\subsubsection{Typical Scales and Units}

We use parameters typical of the Virgo supercluster for illustration.  
Generally the plasma frequency is given by
\begin{equation}
\omega_p^2 = {4\pi\alpha n_e\over m_e} = {n_e\over 10^{-6} {\rm  
cm}^{-3}}
\left(3.7\times 10^{-14} {\rm eV}\right)^2 \ .
\label{eq:plasmafreq}
\end{equation}  
For the supercluster, the plasma density $n_{e} \sim  10^{-6}$ cm$^{-3}$ and  
the magnetic field is of the order of 1 ${\cal B} \sim 0.1 \, \mu$G  
coherent over a distance scale of order 10 Mpc \cite{vallee}.  
As mentioned earlier,  
use of typical ${\cal B}$ values in ${\cal B_{T}}$ formulas
is done to take advantage of the transparent simplicity of that limit.

In the limit of $m_\phi << \omega_{p}$, there is a useful length  
scale $l$ defined by \begin{equation}
l = {2 \omega\over (\omega_p^2 - m_\phi^2)}\approx {\nu\over 10^6\ {\rm  
GHz}}
0.04\ {\rm Mpc}
\label{eq:omega_length}
\end{equation}
where $\nu = {\omega\over 2\pi}$. We refer to this as the oscillation
length in analogy with a similar variable in neutrino physics \cite{Neutrino}. 
The value $10^6\ {\rm GHz} \sim 4  
\ {\rm eV}$ is a handy upper order of magnitude for optical frequencies.

  A typical upper limit \cite{PDG,Rosenberg,raffelt,Brockway}
on the coupling parameter $g_{\phi}$ is
$6\times 10^{-11}$ GeV$^{-1}$.   In Fig. \ref{fig:uniform1} below, we show  
results for $g_{\phi} \sim 10^{-12}$ GeV$^{-1}$, more than one order of  
magnitude smaller than the limit. For a magnetic field of 0.1 $\mu$G we  
find that the product $g_{\phi} {\cal B}$
can be expressed as $g_{\phi}{ \cal B} = 0.215$ Mpc$^{-1}$.  For  
convenience the scales are summarized in different units in Table  
\ref{table:ScalesTable}.

\begin{table}
   \centering

   \begin{tabular}{ccc}
\hline
     quantity & typical value  &  alternate units \\
     \hline
   ${\cal B} $   &   0.1 $\mu$ G    &   $4.78 \times 10^{49 }$ Mpc$^{-2}$   
\\
  $g_{\phi}$  &  $10^{-11} $GeV$^{-1}$  &  $6.4 \times 10^{-50} $ Mpc \\

  $\omega_{p} $ & $ 3.7 \times 10^{-23} $GeV$  \sqrt{\frac{ n_{e}}{  
10^{-6}  cm^{-3}}  }$  & 5.7 $\times 10^{15} \,  \sqrt{\frac{ n_{e}}{  
10^{-6}  cm^{-3}}  }$ Mpc $^{-1}$   \\
  $\omega$ & $10^{-5}$ - 1 \, eV & 1.6 $\times 10^{24}$ - $1.6 \times  
10^{29}$ Mpc$^{-1}$ \\

\hline
\end{tabular}

   \caption{Typical values of dimensionful scales in different units.   
If not otherwise specified we use $\hbar=c=1$.  The value of $g_{\phi}$ 
listed is 
far below published limits of $g_{\phi}< 6 \times 
10^{-11}$ GeV$^{-1}$ \cite{PDG}.  }\label{table:ScalesTable}
\end{table}

\subsubsection{Small Mixing Limit: a Simple Example}

For an example we examine the most innocuous case of mixing angle  
$\theta \ra 0$.  Specifically, this is the limit $$ { g_{\phi}\omega  
{\cal B_T}\over | m_\phi^2 - \omega_p^2 |} <<1. $$ By Eq. \ref{smallg},  
the limit of small mixing is just the same limit in which the naive  
Taylor expansion of Eq. \ref{slop} in small $g_{\phi} \sim 0$ applies.  
In this limit $E_{||}, \, E_{\perp}$ and $\phi$ are the approximate  
propagation eigenstates.  One might think there are no observable  
effects, and in particular, there would be no substantial ``dimming''  
of intensity.

However there is an important relative phase shift in propagation:  \ba  
E_{||}(z) =E_{||}(0) e^{i (\omega+\Delta_{A})z}; \nn \\ E_{\perp}(z)  
=E_{\perp}(0) e^{i  k_{0} z}.  \nn \ea
The physically observable density matrix $\rho$ is given by  \ba &  
\rho&=
\left(\begin{array}{cc} <\,E_{||} E_{||}^{*} \,>   & <\,  
E_{||}E_{\perp}^{*} \,> \\<\,
E_{\perp} E_{||}^{*} \,> &<\, E_{\perp} E_{\perp}^{*}  \,>   
\end{array}\right) , \ea where $<\: >$ denotes the statistical averages  
occurring in propagation. Under coherent conditions this result  
predicts a cumulative rotation of the plane of a linear polarization  
due to the off-diagonal term: \ba  E_{||}E_{\perp}^{*}(z) =  
E_{||}E_{\perp}^{*}(0) e^{  i(   \omega_{p}^{2}/2 \omega    
-\mu_{+}^{2}/2 \omega )z } . \label{examplerot}\ea So long as the  
mixing is small, we may insert the $g_{\phi}^{2}$ expansions of Eq.  
\ref{slop} and predict the angle of rotation increases linearly with  
frequency.  Yet there is always a limit in which the frequency is large  
enough to cause strong mixing.  We explore this for $m_{\phi}^{2}<<  
\omega_{p}^{2}$.

Let the magnitude of the phase angle between $E_{||}$ and $E_{\perp}$  
be denoted $\chi$.  From Eq. \ref{examplerot} we have\ba \chi &=& z (  
\,  \frac{ \omega_{p}^{2} }{2 \omega}   -\frac{ \mu_{+}^{2}}{2 \omega}  
\,)   \sim z \, \frac{ g^{2} {\cal B_{T}}^{2} \omega }{2 \omega_{p}^{2}  
} ,   \:\:\: \omega << {   \omega_{p}^{2} \over g_{\phi}  {\cal B_{T}} };  
\nn \\  \chi &< &  z  g_{\phi}  {\cal B_{T}}. \ea  Using $g_{\phi}  {\cal  
B_{T}} \sim$ Mpc$^{-1}$ as a typical value, the phenomenon of a rotating  
polarization might be readily observed to be linear in $\omega$ at  
radio frequencies for decades above the GHz regime.

Approaching the optical region, both the formula for small mixing and  
the expansion of Eq. \ref{slop} break down, leading to very interesting  
possibilities.

\subsubsection{Comments on Propagation}
It is interesting to contrast the results above with propagation in a
dispersionless and non-dissipative medium, such as ``free-space'' with
gravitational fields.  It is obvious that the intensity of a wave
(Stokes $I$) is preserved up to the kinematic redshift of
propagation \footnote{Reflections, namely the generation of backwards
moving waves, are always possible in any varying medium. We work in the
limit that they are negligible.}.  It is less obvious that the {\it
degree of polarization} (Stokes $Q/I$) cannot be changed by any 2-state  
purely electromagnetic
propagation in a dissipationless medium.  The origin of this is
``unitarity'' of propagation which can be written as a unitary
evolution operator, just as in quantum mechanics \cite{BornWolf}.  As a
consequence of 2-state electromagnetic unitarity, a particular circular  
polarization
(say) can be converted into a linear one, but a linear polarization
cannot be made to come from an unpolarized ensemble. In the weak mixing  
region, an unpolarized density matrix (a multiple of the identity)  
evolves with no change whatsoever. This means that basing observations  
or parameter limits on unpolarized quantities can easily miss effects  
that polarized observables would readily detect.

Spontaneous polarization is more than rare, and in pure electrodynamics  
any form of polarization is usually associated with a corresponding  
extinction. An ordinary polarizer plate is typical, with the degree of  
polarization scaling directly with the
degree of extinction.  The same goes for common sources of polarization  
such as Compton scattering invoked in astrophysics.  Momentarily we  
will discuss stronger mixing cases where mixing of light with  
pseudoscalars can lead to spontaneous polarization.

\subsection{General Density Matrix}

We can now determine the general density (coherency) matrix elements  
after
propagation through distance $z$ in terms of the matrix elements
at the source.  Inasmuch as we can interchange $\vec A $ with $\vec  
E/\omega$, it is convenient to report density matrices of $\vec A$.   
Then:
\begin{eqnarray}
<A^*_\parallel (z) A_\parallel (z)>& = & {1\over 2}
<A^*_\parallel (0) A_\parallel (0)>
\left[1+\cos^2 2\theta
+\sin^22\theta \cos[z(\Delta_\phi - \Delta_A)]\right]  \nonumber\\
&+& {<\phi^*(0)\phi(0)>\over 2}
\left[\sin^2 2\theta
-\sin^22\theta \cos[z(\Delta_\phi - \Delta_A)]\right]  \nonumber\\
&+& \bigg\lbrace{<A_\parallel^*(0)\phi(0)>\over 2}
\bigg[\sin 2\theta\cos 2\theta - \sin 2 \theta \cos 2\theta
\cos[z(\Delta_\phi - \Delta_A)]\nonumber\\
&-& i\sin 2\theta\sin[z(\Delta_\phi - \Delta_A)]\bigg] +  
c.c.\bigg\rbrace
\end{eqnarray}

We gave the most general expression above keeping all the correlators
at $z=0$.  One might think it reasonable to assume that   
$<A_\parallel^*(0)\phi(0)>$ is zero.  However in considering  
propagation through an intergalactic medium one expects a large number  
of magnetic domains uncorrelated
with one another.  The general expression is needed to describe  
sequentially the propagation through different domains.  Given the  
randomness of such processes, the generic situation is one where all  
the correlators on the right hand side will
be nonzero.

We also find,
\begin{eqnarray}
<A^*_\parallel (z) A_\perp(z)> &=&(\cos^2\theta e^{iF z}
  + \sin^2\theta e^{iG z})
<A_\parallel^*(0) A_\perp(0)>\nonumber\\
&+& \cos\theta\sin\theta (e^{iF z}
-  e^{iG z}) <\phi^*(0) A_\perp(0)>
\end{eqnarray}
and $<A_\perp^*(z)A_\perp(z)> = <A_\perp^*(0)A_\perp(0)>$.
The phase factors $F$ and $G$ are given by
$$ F = \omega\sqrt{1-{\omega_p^2\over \omega^2}}
-\omega\sqrt{1-{\mu_+ \over \omega^2}}
\approx {1\over 2\omega}
(\mu_+ -\omega_p^2) $$
$$ G = \omega\sqrt{1-{\omega_p^2\over \omega^2}}
-\omega\sqrt{1-{\mu_- \over \omega^2}}
\approx {1\over 2\omega}
(\mu_- -\omega_p^2) $$
Here again it is reasonable to expand $F$ and $G$ and keep only
the leading order terms in $ \omega_p^2/ \omega^2$ and $\mu_\pm/  
\omega^2$.
Using these correlation functions we can compute reduced Stokes
parameters $I,Q,U,V$ (or $S_0,S_1,S_2,S_3$)
\begin{eqnarray}
I & =& <A_\parallel^*(z)A_\parallel(z)> +  
<A_\perp^*(z)A_\perp(z)>\nonumber\\
Q &=& <A_\parallel^*(z)A_\parallel(z)> -  
<A_\perp^*(z)A_\perp(z)>\nonumber\\
U &=& <A_\parallel^*(z)A_\perp(z)> +  
<A_\perp^*(z)A_\parallel(z)>\nonumber\\
V &=& i(-<A_\parallel^*(z)A_\perp(z)> + <A_\perp^*(z)A_\parallel(z)>
\label{stokes}
\end{eqnarray}
at any position $z$.  The standard Stokes parameters of the same name  
are simply $\omega^{2}$ times the above.  We will often remove the  
scale of intensity, normalizing $I=2$, as in an unpolarized wave  
$|A_{||}|^{2}= |A_{\perp}|^{2}=1. $

\subsubsection{Special Cases 1: Spontaneous Appearance of Polarization}

Suppose the initial beam is unpolarized and the initial correlator
$<A_\parallel^*(0)\phi(0)>$ is also zero. 
We scale out normalizations so that 
$$<A_\parallel^*(0)A_\parallel(0)> = <A_\perp^*(0)A_\perp(0)> = 1$$
and assume that 
all other correlators vanish at $z=0$. The expressions simplify to:
\begin{eqnarray}
<A^*_\parallel (z) A_\parallel (z)>& = & {1\over 2}\left(1+\cos^2  
2\theta
+ \cos[z(\Delta_\phi-\Delta_A)]\sin^2 2\theta\right) +{1\over  
2}\nonumber\\
&\times &  <\phi^*(0)\phi(0)>
(1-\cos[z(\Delta_\phi-\Delta_A)] )\sin^2 2\theta\\
<A^*_\parallel (z) A_\perp(z)> &=&0 \\
<A^*_\perp (z) A_\perp(z)> &=&1
\end{eqnarray}
\begin{figure}
\hskip -0.6in
\psfig{file=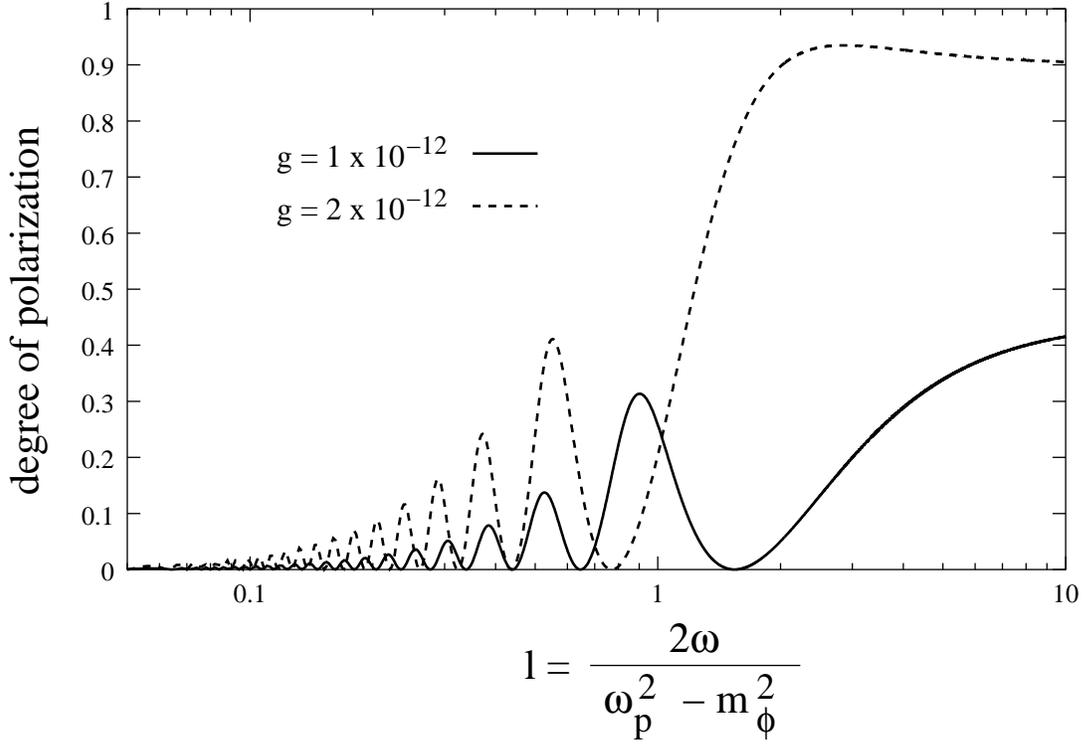}
\caption{The degree of polarization $p$ as a function of the length
parameter $l$ (Eq. \ref{eq:omega_length})
for uniform background propagation
for two different value of the coupling parameter $g_\phi$. Values of
$g_{\phi}$ are given in units of GeV$^{-1}$;  ${\cal B}=0.1\ \mu$ G,   
$n_e = 10^{-6}$ cm$^{-3}$. The length scale $l$ is in units of Mpc   
($\nu/ 10^6$ GHz). The propagation distance is taken to be 10 Mpc.}
\label{fig:uniform1}
\end{figure}
This wave spontaneously acquires a linear
polarization oriented along $\vec {\cal B}$ during propagation.  No  
circular polarization is developed.  We plot the degree of polarization  
in Fig. \ref{fig:uniform1} for initial conditions $<\phi^*(0)\phi(0)> =  
0$. The degree of
polarization $p$ here is given by
\begin{equation}
p = {|Q|\over I} = {|<A^*_\parallel (z) A_\parallel (z)> -
<A^*_\perp (z) A_\perp(z)>|\over <A^*_\parallel (z) A_\parallel (z)>
+ <A^*_\perp (z) A_\perp(z)>}\ ,
\end{equation}
given Stokes $U=V=0$.
The degree of polarization accumulates to a sizeable magnitude and  
could produce observable consequences even for exceedingly small  
couplings.
In Fig. \ref{fig:uniform1} the degree of polarization is shown for two 
different values of the coupling parameter $g_\phi = 1\times 10^{-12},\ 
2\times 10^{-12}$ GeV$^{-1}$. The background medium parameters are chosen
to be ${\cal B} = 0.1\ \mu G$ and $n_e=10^{-6}$ cm$^{-3}$. The propagation
distance is taken to be 10 Mpc.
This is also the distance over which the background magnetic field is assumed
to be coherent.

The experimental signature of this mixing for a uniform background  
would require fitting the observed degree of polarization to a number  
of parameters.  As shown by the previous examples, the extrapolation  
between radio frequency and optical frequency observations needs care  
and attention to series expansions.  Nevertheless the spontaneous  
polarization due to mixing is clearly distinguishable from extinction,  
which is expected to show a simple monotonic increase in the
degree of polarization with frequency.

\begin{figure}
\psfig{file=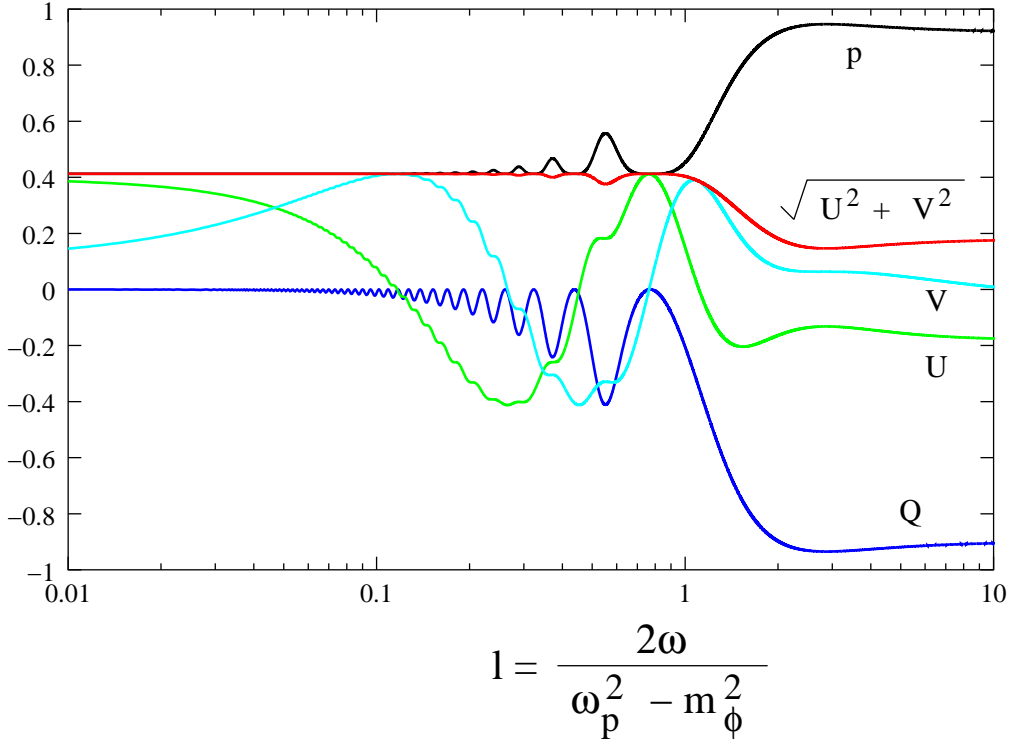}
\caption{Reduced Stokes parameter ratios $Q/I,U/I$ and $V/I$ as a  
function of the length
parameter $l$ in units of Mpc, (Eq. \ref{eq:omega_length}),
for uniform background propagation with $g_\phi=2\times 10^{-12}$ GeV$^{-1}$.
The degree of polarization $p=\sqrt{Q^2+U^2+V^2}/I$
and the variable $\sqrt{U^2 + V^2}/I$ are also shown.  Other parameters  
are ${\cal B}=0.1\ \mu$ G, $n_{e} =10^{-6} cm^{-3}$, Propagation distance  
 =10 Mpc, $m_\phi/\omega_p=0.1$. The initial state of the  
polarization has been chosen randomly such that $Q/I = 0$, $U/I = 0.4$
and $V/I = 0.1$. 
}
\label{fig:stokes1}
\end{figure}

\subsubsection{Special Cases 2: Initial Polarization}

Consider now the case where the incident beam is partially
polarized. A sample of results for randomly chosen intial conditions is
shown in Fig. \ref{fig:stokes1}. The Stokes
$Q/I$, $U/I$ and $V/I$
parameters are presented as a function of the length parameter $l$,  
equivalent to the frequency of the incident wave.  Initial conditions  
are $<\phi^*(0)\phi(0)>=0$, $ <A^*_\parallel(0)\phi(0)>=0$ and $  
<A^*_\perp(0)\phi(0)>=0$.  A more general result is discussed later.  
One interesting pattern that emerges from this figure is
that $U$ and $V$ show a much larger variation with frequency compared to
to $Q$. This is related to observations of Ref. \cite{PJ} that at low  
frequencies the dominant physical effect arises due to the phase in  
$A_\parallel$, an
evolution of the polarization direction of the wave, which is permitted  
by 2-state unitarity.
The figure supports near conservation of $\sqrt{U^2 + V^2}$ which
is relatively independent of frequency, both in the low and high
frequency regime. Indeed this variable shows spectral dependence only
in a relatively narrow region. In the high frequency regime, of course,
all the Stokes parameters show dependence on frequency.

\begin{figure}
\hskip -1.0in
\psfig{file=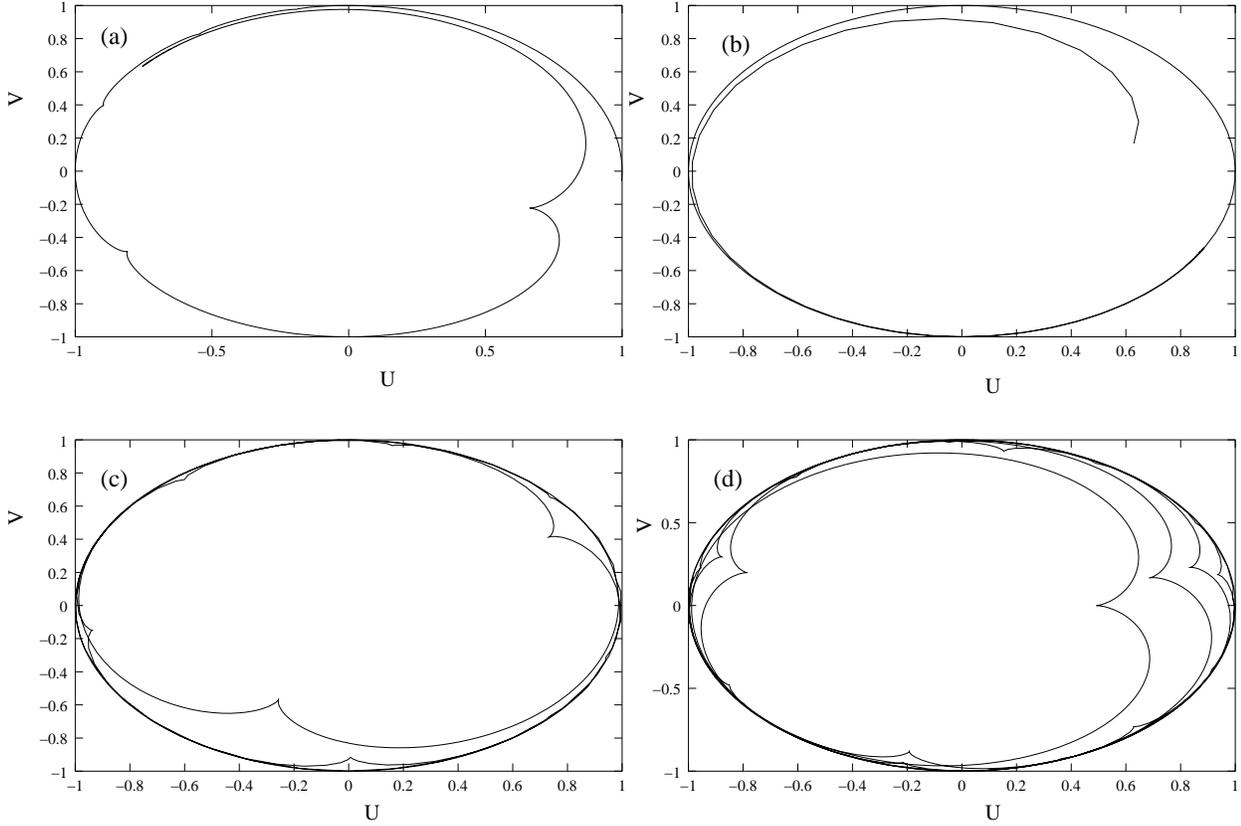}
\caption{A sample of results showing the correlation between the Stokes
parameters $U$ and $V$ for randomly chosen parameters and initial
state of polarization under propagation in a
uniform background. Parameters (in arbitrary units) are
(a) $g_{\phi} {\cal B} = 2,\ L = 10,\  0.04<l<20$,
(b) $g_{\phi} {\cal B} = 10,\ L = 10,\  0.4<l<300$,
(c) $g_{\phi} {\cal B} = 1,\ L = 50,\  0.2<l<800$ and
(d) $g_{\phi} {\cal B} = 10,\ L = 10,\  0.04<l<100$;    
$m^2_\phi/\omega^2_p=0.1$
for all the plots. A simple correlation is seen only for frequencies  
larger than
a minimum frequency. At low frequencies the relationship becomes
very complicated.
  }
\label{fig:cor_uniform}
\end{figure}

\begin{figure}
\hskip -1.0in
\psfig{file=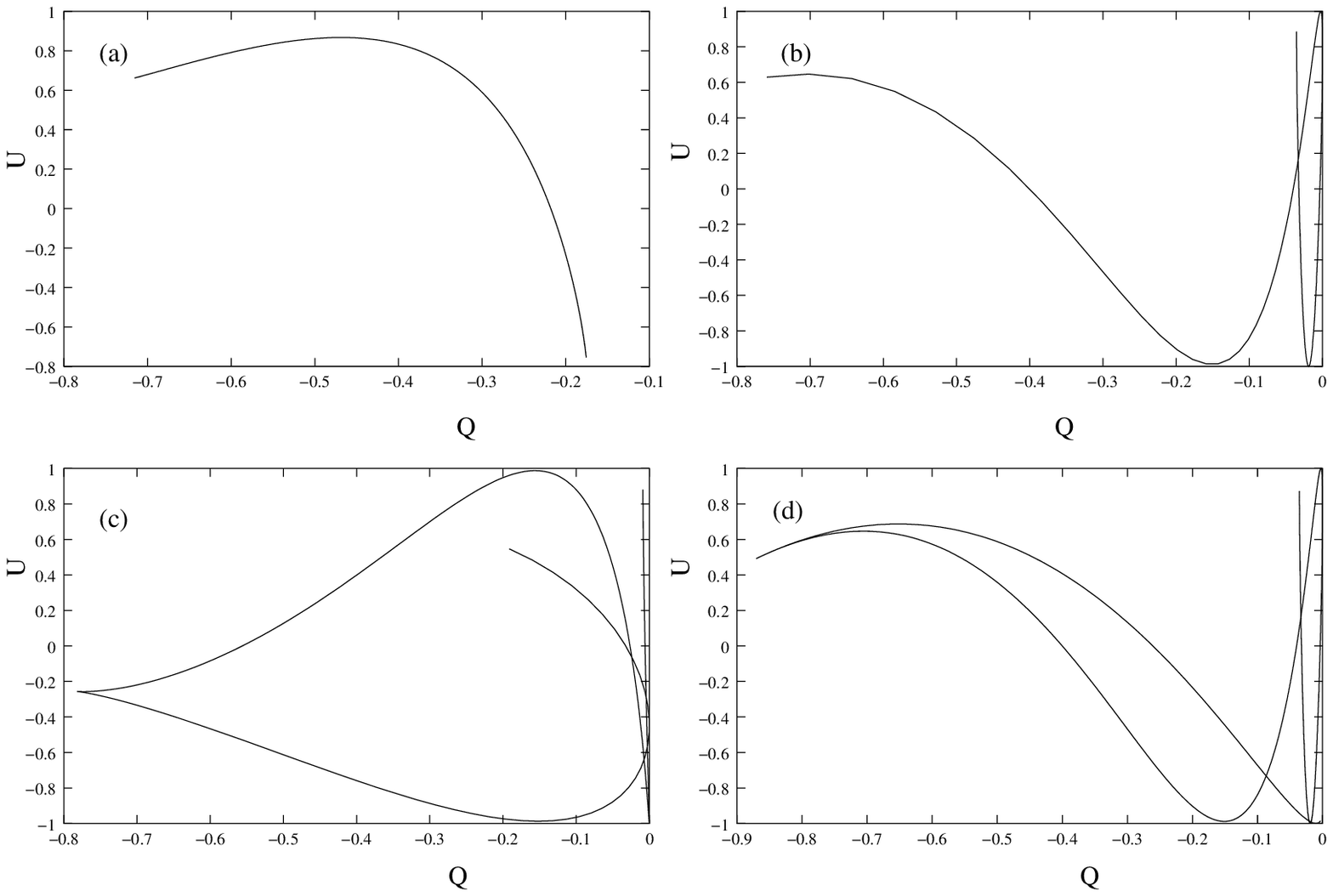}
\caption{A sample of results showing the correlation between the Stokes
parameters $Q$ and $U$. Parameters and initial
states of polarization are taken to be same as used in Fig.
\ref{fig:cor_uniform}. The oscillation length $l$ is
(a) $ 1.1<l<20$,
(b) $0.4<l<300$,
(c) $ 1.7<l<800$ and
(d) $ 0.27<l<100$.
  The lower limits on $l$ are taken slightly larger than the values used
in  Fig. \ref{fig:cor_uniform} in order that a simple relationship
  emerges between $Q$ and $U$.
  }
\label{fig:QU_uniform}
\end{figure}

A possible signal of the pseudoscalar mixing is the correlation
between the different Stokes parameters, in particular between
$U$ and $V$. In Fig. \ref{fig:cor_uniform} we show a sample of
results to illustrate the correlation between the Stokes parameters
$U$ and $V$ for some randomly chosen parameters and initial
state of polarization. We find these Stokes parameters do show an
observable dependence on each other. The dependence of $U$ on $V$ is
found to approximately follow an ellipse for large frequencies.
  If the frequency is small such that the oscillation length $l<<L$, where  
$L$ is the propagation distance, then we find that the dependence of  
$U$ on $V$ is considerably more
complicated. Remarkably, the Stokes parameter $Q$ (degree of  
polazarization) also shows a simple dependence on $U$ and $V$  
(polarization direction controllers) at large frequencies. This is  
shown for some randomly
chosen parameters in Fig. \ref{fig:QU_uniform}.

These illustrative calculations show that a fascinating range of  
physical phenomena are generated by the mixing of light and  
pseudoscalars.

\section{Pseudoscalar-Photon Mixing: Slowly Varying Background}

For the rest of the paper we will consider cases in where the
background is slowly varying.
There exists a large literature regarding the integration of
differential equations with slowly varying parameters. The reader
interested in the mathematical problem can consult the classic
reference of Olver \cite{Olver}.  It is an understatement to say that 
the variety of situations
is huge.  Rather than make sweeping statements, we will use a 
workmanlike series of definitions and approximations that we check are 
suited to the limits we study.

First restrict attention to the simple case where the
direction of $\vec{\cal B}$ is uniform. Assume
that the background varies sufficiently slowly to justify neglect of
terms involving the derivatives
of the background plasma density (or the magnitude of the magnetic
field).  We thus enter into
 the {\it adiabatic approximation}, a general branch of mechanics,
which holds that the amplitude in an eigenstate is relatively constant
when the parameters defining the eigenstate change sufficiently slowly. 
  We again obtain two decoupled equations Eq. \ref{eq:diagonalized}
  which can be solved by using the ansatz
  \begin{eqnarray}
  \overline {A}_\parallel(z) &=& \overline{A}_\parallel(0)
  e^{i\omega z+i\int_0^z \Delta_A dz'}\nonumber\\
  \overline {\phi}(z) &=& \overline{\phi}(0) 
  e^{i\omega z+i\int_0^z \Delta_\phi
  dz'}
  \end{eqnarray}
The phase factors are given in Eq. \ref{eq:phases}.

In the case there is a conservation law -- the conservation of
transmitted energy flux, similar to the conservation of probability in
quantum mechanics - the WKB approximation would indicate
certain pre-factors of order $1/\sqrt{k(z)}$, where $k(z)$ are the
exponential integrands.  We review this with an iterative WKB procedure 
developed in Section 5
which controls coefficients in a certain expansion.   However extensive 
calculations support wide applicability of the simpler formulas, 
inasmuch as $k(z)$ tends to cancel out of polarization ratios and not 
to develop cumulative run-outs in many circumstances.


Using these results we can now evaluate the electromagnetic field at any
position. For this purpose we express the fields $\overline A(0)$ and
$\overline\phi(0)$ in terms of $A_\parallel(0)$ and $\phi(0)$ and
the mixing angle $\theta(0)$. The final expressions for the fields
$A_\parallel(z)$ and $\phi(z)$ are given by
\begin{eqnarray}
A_\parallel(z) & = & e^{i\omega z}A_\parallel(0) \left[\cos\theta\cos\theta_0
\exp\left(i\int_0^z\Delta_A dz'\right)+\sin\theta\sin\theta_0
\exp\left(i\int_0^z\Delta_\phi dz'\right) \right]
\nonumber\\
& + & e^{i\omega z}\phi(0) \left[\cos\theta\sin\theta_0
\exp\left(i\int_0^z\Delta_A dz'\right)-\sin\theta\cos\theta_0
\exp\left(i\int_0^z\Delta_\phi dz'\right) \right]
\nonumber\\
\phi(z) & = & e^{i\omega z}A_\parallel(0) \left[\sin\theta\cos\theta_0
\exp\left(i\int_0^z\Delta_A dz'\right)-\cos\theta\sin\theta_0
\exp\left({i\int_0^z\Delta_\phi dz'}\right) \right]
\nonumber\\
& + & e^{i\omega z}\phi(0) \left[\sin\theta\sin\theta_0
\exp\left(i{\int_0^z\Delta_A dz'}\right)+\cos\theta\cos\theta_0
\exp\left(i{\int_0^z\Delta_\phi dz'}\right) \right]
\nonumber\\
& &
\end{eqnarray}
where on the right hand side $\theta = \theta(z)$ and $\theta_0 =  
\theta(0)$.
The resulting expressions for the coherency matrix elements is given by
\begin{eqnarray}
<A^*_\parallel(z)A_\parallel(z)> &=& {<A^*_\parallel(0)A_\parallel(0)>
\over 2}
\left[1 + \cos 2\theta\cos 2\theta_0 + \sin 2\theta\sin  
2\theta_0 \cos\Phi\right]\nonumber\\
&+& {<\phi^*(0)\phi(0)>\over 2}
\left[1 - \cos 2\theta\cos 2\theta_0 - \sin 2\theta\sin  
2\theta_0
\cos\Phi \right]\nonumber\\
&+& \bigg( {<A^*_\parallel(0)\phi(0)>\over 2}
 \bigg[\cos 2\theta\sin 2\theta_0 - \sin 2\theta\cos 2\theta_0
  \cos\Phi \nonumber\\
&+& i\sin 2\theta\sin\Phi
\bigg] + c.c.\bigg)
\label{eq:M11}
\end{eqnarray}
where
\begin{equation}
\Phi = \int_0^z(\Delta_A-\Delta_\phi) dz'
\end{equation}

\begin{eqnarray}
<A^*_\parallel(z)A_\perp(z)> &=& <A^*_\parallel(0)A_\perp(0)>
\bigg[\cos\theta\cos\theta_0 \exp\left({i\over 2\omega}
\int_0^z dz'[\mu^2_+-\omega_p^2]\right) \nonumber\\
&+& \sin\theta\sin\theta_0 \exp\left({i\over 2\omega}
\int_0^z dz'[\mu^2_--\omega_p^2]\right)\bigg]
\nonumber\\
&+& <\phi^*(0)A_\perp(0)>
\bigg[\cos\theta\sin\theta_0 \exp\left({i\over 2\omega}
\int_0^z dz'[\mu^2_+-\omega_p^2]\right) \nonumber\\
&-& \sin\theta\cos\theta_0 \exp\left({i\over 2\omega}
\int_0^z dz'[\mu^2_--\omega_p^2]\right)\bigg]\ .
\label{eq:M12}
\end{eqnarray}
The above result is valid as long as the medium changes slowly so that
the term
\begin{equation}
2O^T(\partial_z O)\partial_z\left( \matrix{\overline A\cr \overline  
\phi}
\right) = 2i\omega\theta'\left(\matrix{0 & -1\cr 1 & 0} \right)
\left(\matrix{\overline A\cr \overline \phi} \right)
\end{equation}
is negligible compared to mass term
$$   \left(\matrix{\mu^2_+ & 0\cr 0 &\mu^2_- }\right)
\left(\matrix{\overline A\cr \overline \phi}  \right)\ .$$
We are justified in ignoring the derivative of
the mixing angle $\theta $ if
\begin{equation}
|\theta'| << {\mu^2_\pm\over 2\omega}
\label{eq:adiabatic}
\end{equation}
The condition for adiabaticity will be formulated in greater detail
in section 5.  There we estimate the transition probability from
one local eigenstate to another.  As we shall see, the condition given
above, Eq. \ref{eq:adiabatic}, is sufficient provided the difference
between $\mu_+$ and $\mu_-$ is of the order of these eigenvalues.
If there is a delicate cancellation between these two eigenvalues then
on the right hand side in Eq. \ref{eq:adiabatic} more care and  
specialized techniques may be needed.

\subsection{Resonance}

We next discuss the interesting
case where the plasma frequency $\omega_p$ becomes equal to the pseudoscalar 
mass $m_\phi$ somewhere along the path. 
Given the wide range of plasma frequencies observed in the astrophysics
there may be many situations where such resonance might occur. Examples
include propagation of light or pseudoscalars from a dense medium, such
as an AGN, GRB or cluster of galaxies 
into the intergalactic medium. Another example includes propagation of 
light from pulsar magnetosphere into the interstellar medium. 
Let the radiation initially
propagate from the region $z=0$ containing high plasma density  
$\omega_p>m_\phi$ towards increasing $z$ where a resonance occurs.
Assume that initially the mixing angle $\theta <<1$,  and recall
\ba  \tan 2\theta = \frac{ 2g_{\phi}\omega {\cal B_T} }{  
m_\phi^2 - \omega_p^2} <<1 \label{eq:tan2theta2}. \ea   Let
the plasma density decrease slowly along the path such that at the  
observation
point $m_\phi > \omega_p$.  Somewhere along the path at $z=z^{*}$ the  
right
hand side of Eq. \ref{eq:tan2theta2} becomes infinite, which we will  
implement with angle $2\theta(z_{*}) \ra  \pi/2$ . To finally fix the  
initial conditions, let $<A^*_\parallel(0)A_\parallel(0)> \neq 0 $,  
$<A^*_\parallel(0)A_\perp(0)>\neq 0 $
and $<A^*_\perp(0)A_\perp(0)>\neq 0 $ , that is, a generically mixed  
wave. The remaining correlators, which involve $\phi$, are taken to be 
zero at $z=0$.  We compare the conditions after traveling a distance $\Delta z=L$ 
to the observation point.  From  Eq. \ref{eq:M11}, \ref{eq:M12} we find: 
\begin{eqnarray}
<A^*_\parallel(L)A_\parallel(L)> &\approx& 0\ ,\nonumber\\
<A^*_\parallel(L)A_\perp(L)> &\approx& 0\ ,\nonumber\\
<A^*_\perp(L)A_\perp(L)> &=& <A^*_\perp(0)A_\perp(0)>
\label{eq:reso_coherency}
\end{eqnarray}  
if $|\tan 2\theta|<<1$ at $z=L$. 
These are predicted independent of frequency and the  
distance travelled, provided the wave crosses the region where  
$m_\phi=\omega_p $. Almost independent of the state of polarization at  
the origin,
the wave becomes completely linearly polarized at the observation
point.  The orientation of the electric field vector is perpendicular
to the background magnetic field. \footnote{Generally all resonant  
systems have a special phase at resonance.  We acknowledge R. Buniy for  
discussions and work long ago on similar resonant phase evolution in  
classical mechanics.}

Studying resonance using a slowly varying approximation requires some  
care.   The consistent requirement
is that the assumption of slowly varying background be satisfied. We  
now make
this assumption more precise in order to obtain a constraint on the
derivative of the background plasma frequency.  For this purpose we
assume that the background plasma density varies linearly with position
such that
\begin{equation}
{m_\phi^2\over \omega_0} - {\omega_p^2\over \omega_0} = C-\alpha z
\label{eq:linear}
\end{equation}
where $C$ and $\alpha$ are constants, $z$ is the distance of  
propagation and $\omega_0$ is some chosen value of the frequency of the wave.
It is convenient to work with the rescaled variables, 
\be \tilde\omega_p^2={\omega_p^2\over\omega_0}\ \  ,\  
\tilde m_\phi^2={m_\phi^2\over \omega_0} \ \ ,\ \tilde \omega={\omega\over\omega_0}\ , 
\label{eq:rescaled}\ee 
since for cosmological
applications $\tilde\omega_p^2$ 
has a value of order unity in units of Mpc$^{-1}$. 
The origin of the wave corresponds to $z=0$ and the observation point
$z=L$. 

If at $z=0$ the plasma frequency is larger than pseudoscalar
mass then the constants $C$ and $\alpha$ are taken to be negative.
We assume that the wave crosses the resonance region where $m_\phi
=\omega_p$. As discussed earlier, in this case our approximation is
valid if
\begin{equation}
|\theta'| << {|\mu^2_+-\mu^2_-|\over 2\omega}\ .
\label{eq:adiabatic1}
\end{equation}
This is discussed in greater detail in the next section. This constraint
reduces to
\begin{equation}
|\alpha| << 4(g_{\phi} {\cal B})^2\tilde\omega
\label{eq:adiabatic2}
\end{equation}
at the resonance point $m_\phi=\omega_p(z)$.  Now we come to the point  
of these numbers: for the coupling $g_\phi=6\times 10^{-11}$ GeV$^{-1}$ the  
constraint is satisfied
for the magnetic field, plasma density 
and the distance scale corresponding to the
Virgo supercluster, provided the frequency of the wave $\omega >> 1.7\times
10^{-4}$ eV. Since this is well below 
optical frequencies and well above the regime of GHz radio frequency 
studies, one should be prepared to see different phenomena in the 
optical and radio regimes. One  
can both use a slowly varying approximation, and observe resonant  
mixing consistently.

Let us turn to the phases that appear in Eq. \ref{eq:M12} for the
case of linearly varying background. We find
\begin{eqnarray}
\int_0^z dz'\ {\mu^2_\pm - \omega_p^2\over \omega_0} &=& {1\over 4}[-\alpha z^2
+2zC] \pm {C\over 4\alpha}\sqrt{4g^2{\cal B}^2\tilde\omega^2 + C^2}\nonumber\\
&\pm&{\alpha z-C\over 4\alpha}\sqrt{4g^2{\cal B}^2\tilde\omega^2 + 
(\alpha z-C)^2} \nonumber\\
&\pm& {(g_{\phi} {\cal B}\tilde\omega)^2\over \alpha}
\Bigg[\log\Bigg(\alpha z-C+\sqrt{(2g_{\phi} {\cal B}\tilde\omega)^2+(\alpha  
z-C)^2}\ \Bigg)
\nonumber\\
&-& \log\Bigg( -C + \sqrt{(2g_{\phi} {\cal B}\tilde\omega)^2 +C^2} \ \Bigg)
\Bigg]
\label{eq:lambdapm}
\end{eqnarray}
Since these phases are equal to
$\int_0^z dz'[\mu^2_\pm - \omega_p^2]/2\omega$
we find that for a wide range of parameter space these phases
vary roughly as $1/\omega$. The corresponding polarization observables  
change roughly inversely proportional to $\omega$.

We can understand this behaviour of the phases appearing in
the correlator $<A_\parallel^*(z) A_\perp(z)>$ as follows. Consider the  
phase
$\int_0^z dz' (\mu^2_+-\omega_p^2)/2\omega$, which
is basically the integral of the difference of the squared masses of
one of the eigenmodes
and the perpendicular component divided by $2\omega$, as expected
for relativistic particles. In the limit of small mixing the difference
of the squared masses is proportional to the mixing angle squared and
hence is proportional to $\omega^2$. In this limit the phase increases
linearly with $\omega$. However if $\omega_p^{2} -m_\phi^{2} \ra 0 $
along the path there is a limit interchange.   We need a different  
series expansion, for example \ba    \mu_{+}^{2} \sim   g_{\phi}{\cal  
B_{T}}\omega +\frac{1}{2}(\omega_{p}^{2}+ m_{\phi}^{2}) + \frac{   
(\omega_{p}^{2 } - m_{\phi}^{2 })^{2 }  }{8  g_{\phi}{\cal  
B_{T}}\omega}.   \ea   This contradicts perturbation theory in  
$g_{\phi}$, because resonance is always a non-perturbative phenomenon,  
as signaled by the appearance of $1/g_{\phi}$ in the expansion.  In any  
event, given that $\mu_{+}^{2}/\omega$ occurs, the linear increase of  
phase flattens out near $\omega_{p}^{2 } - m_{\phi}^{2 } \sim 0$, and  
can even decrease for large $\omega$.

So far we made the resonance occur.  In the limit of small frequencies  
the phase factors become very large.  For example in Eq.  
\ref{eq:lambdapm} the first factor contributes
the term $m_\phi^2z/2\omega$ to the phase.
  For large propagation distance
$z$ and small $\omega$ this factor is clearly much greater than
unity. Hence one expects that in general the Stokes parameters will show
rapid fluctuations as a function of $\omega$ in this limit. However  
since $g_{\phi} {\cal B}$ occur in a product, the small $g_{\phi}$  
limit can again be reversed if ${\cal B}$ becomes very large!

Continuing, we evaluate the phase $\Phi$ appearing in Eq. \ref{eq:M11} for  
the
case of a linearly varying background plasma frequency, Eq.  
\ref{eq:linear}.
Results can be extracted from the integrals given in Eq.  
\ref{eq:lambdapm}
by using
\begin{equation}
\Phi = \int_0^z (\Delta_A-\Delta_\phi)dz' = \int_0^z 
\left({\mu^2_-\over 2\omega} -{\mu^2_+ \over 2\omega  }\right) dz' \ .
\end{equation}

\begin{figure}
\psfig{file=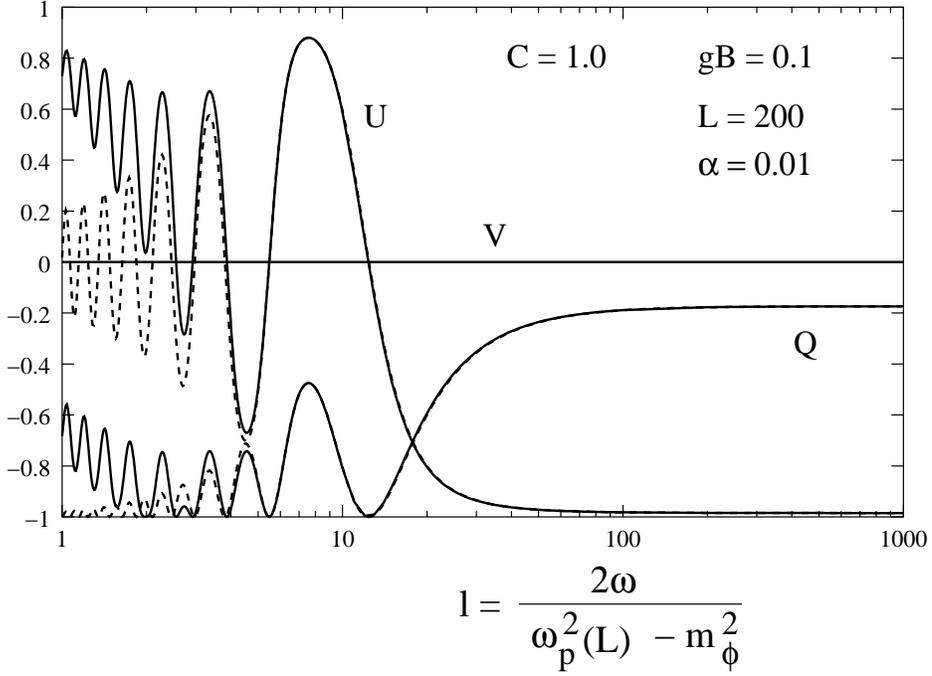}
\caption{The Stokes parameters $Q/I, \, U/I$ and $V/I$ as a function of  
the length
parameter $2\omega/(\omega_p^2(z)-m_\phi^2)$ (in arbitrary units)
for the case of resonant mixing. Solid and dashed lines represent the  
results obtained
by direct numerical integration and by using the analytic expressions  
Eqs. \ref{eq:M11} and
\ref{eq:M12} in the adiabatic
approximation.  The analytic result is in
very good agreement with numerical result as long as Eq.  
\ref{eq:adiabatic2}
is satisfied.  Here  $\tilde\omega_p^2(z)
= \tilde m_\phi^2-C+\alpha z$;  $g_{\phi} {\cal B} =0.1$; $\tilde m_\phi^2=1$. 
Initial conditions at  
$z=0$ are $p=1.0$, $Q=0.0$, $U=1.0$ and $V=0.0$.
}
\label{fig:resonant1}
\end{figure}

\begin{figure}
\psfig{file=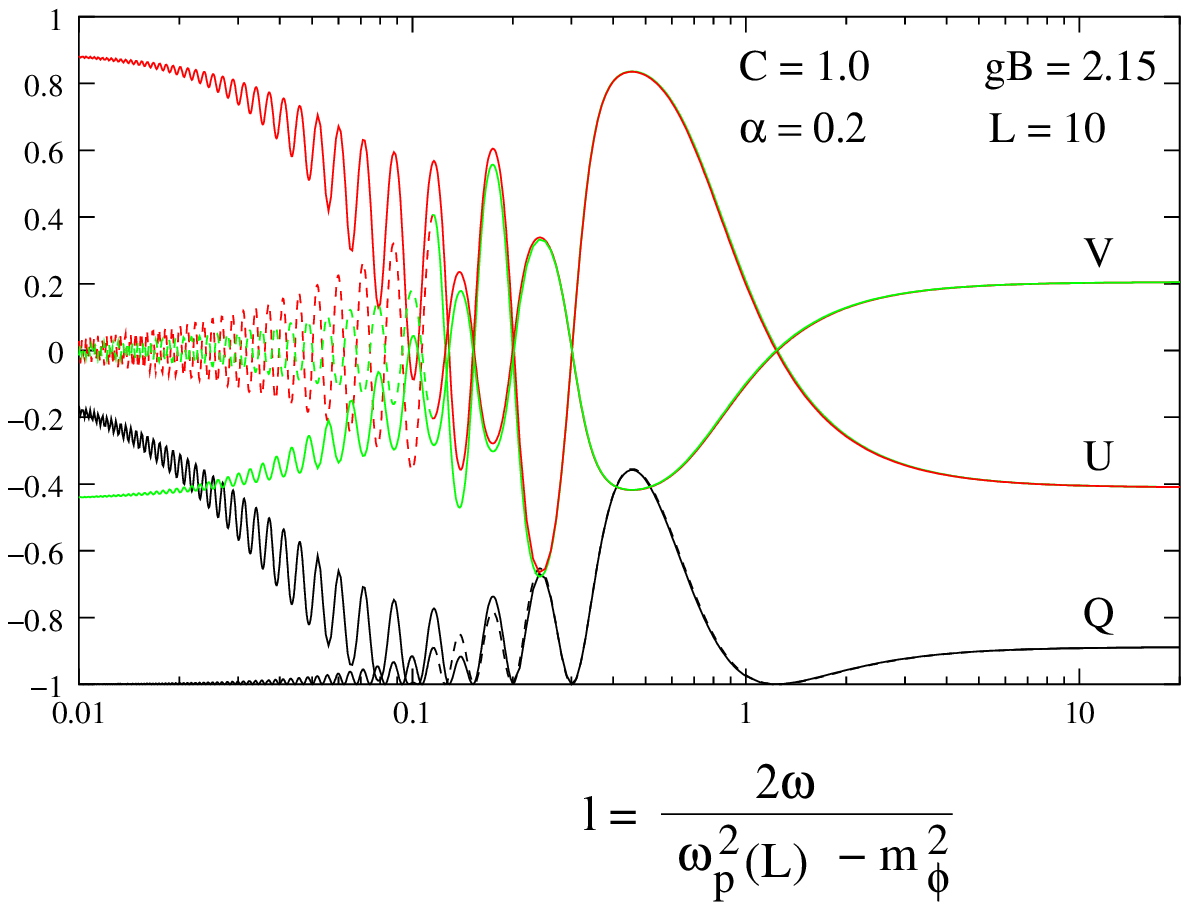}
\caption{Another example of Stokes parameters for resonant
mixing. Parameters $g_{\phi} {\cal B}$ and $L$
are approximately the same as Fig. \ref{fig:stokes1} and $\tilde m_\phi^2=1$.
Initial conditions are $p=1.0$, $Q=0.0$, $U= 0.894$ and $V=-0.447$.
Solid and dashed lines represent the results obtained
by direct numerical
integration and by using the analytic expressions Eqs. \ref{eq:M11} and
\ref{eq:M12} in the adiabatic
approximation. The analytic result is in
very good agreement with numerical result as long as Eq.  
\ref{eq:adiabatic2}
is satisfied. }
\label{fig:resonant2}
\end{figure}

In fig. \ref{fig:resonant1} and \ref{fig:resonant2}
we show two examples of the result expected
in the case of resonant mixing. Here the initial state of polarization
has been chosen randomly. The parameters values  (in arbitrary
units) given in the figure may be translated to those relevant for cosmological 
propagation by taking the units in the appropriate powers of Mpc. 
The pseudoscalar
mass is chosen such that $\tilde m_\phi^2 = m_\phi^2/\omega_0=1$, where
$\omega_0$ is defined in Eq. \ref{eq:linear}.
As expected, observable effects are
large
at low frequencies compared to the corresponding results of a uniform
background. We also find rapid fluctuations in this
region. The analytic result, obtained in the adiabatic limit, is found  
to
be in good agreement with the numerical result as long as Eq.
\ref{eq:adiabatic2} is satisfied. At smaller frequencies the two results
start to disagree. As expected, we numerically find that the
Stokes parameters approach their initial values in the limit  
$\omega\rightarrow 0$.

\begin{figure}
\psfig{file=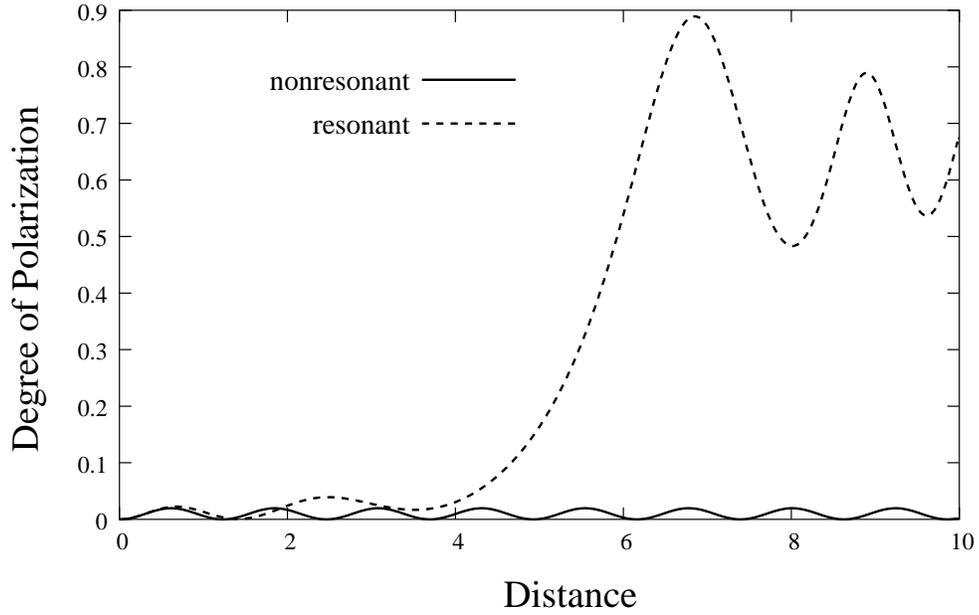}
\caption{Degree of polarization as a function of distance for 
the case of resonant mixing. The wave is taken to be unpolarized
at source. The parameters: $g_\phi{\cal B} = 1.0$, $\tilde m^2_\phi = 1$, 
$C = -0.5$, $\alpha = -0.1$, $l=2\omega/(\omega_p^2(L) - m_\phi^2)
= 0.2$. The curve for uniform background (nonresonant mixing)
is shown for comparison. 
All the parameters in this case 
are same as above, with the (uniform) mixing
length $l=2\omega/(\omega_p^2 - m_\phi^2) = 0.2$.  }
\label{fig:distance}
\end{figure}

In fig. \ref{fig:distance} we show the dependence of the degree of 
polarization on the distance of propagation for the case of
resonant mixing. Here the wave is taken to be unpolarized at source. 
The remaining parameters are taken to be $g_\phi{\cal B} = 1.0$, 
$m_\phi = 1$, $C = -0.5$, $\alpha = -0.1$ and the oscillation length
at the observation point $l=2\omega/(\omega_p^2(L) - m_\phi^2)
= 0.2$. The figure clearly shows the large change in the degree 
of polarization as the wave crosses the resonant point.
For comparison we also show the case of uniform background with
essentially the same parameters as for the resonant case. The 
only difference is that the (uniform) oscillation length parameter
$l = 2\omega/(\omega_p^2 - m_\phi^2)=0.2$. In this case the
wave acquires a very small degree of polarization in comparison 
to the resonant case.

\section{Adiabatic Approximation and Corrections}
In this Section we analyze the pseudoscalar-photon mixing
for slowly varying background more systematically. This will give us an
estimate of corrections to our results above.  We assume the {\it 
adiabatic limit} applies: for any changes of parameters, we take 
derivatives such as $\p {\cal B} /\p z \ra 0$, with the propagation 
distance $L \ra \infty$, and the cumulative change $\Delta {\cal B} $ 
fixed.

Borrowing standard notation from quantum mechanics,
we rewrite the basic equation,
Eq. \ref{eq:mixing}, as follows
\begin{equation}
(\omega^2+\partial_z^2)|\psi> - M|\psi> = 0
\label{eq:waveequation}
\end{equation}
We denote the eigenvectors of the mixing matrix as $|n>$, where
in the case of resonant mixing of $A_\parallel$ with $\phi$,
$n=1,2$. The eigenvalue equation can be written
as
\begin{equation}
M|n> = \mu^2_n|n>
\label{eq:eigenvalueeq}
\end{equation}
Here the eigenvectors also vary slowly with $z$.
The general solution to Eq. \ref{eq:waveequation} can be written
as
\begin{equation}
|\psi> = \sum_n a_n(z) e^{i\int_0^z dz'\omega_n}|n>
\end{equation}
where $\omega_n \approx \omega-{\mu_n^2\over 2\omega}$
Substituting this into Eq. \ref{eq:waveequation}, and taking the overlap
of the resulting equation with $<m|$ we find
\begin{equation}
\partial_z a_m = -{1\over 2}{\partial_z \omega_m\over
\omega_m}a_m - {1\over \omega_m}\sum_n a_n\omega_n
e^{i\int_0^z dz'(\omega_n-\omega_m)}
<m|\partial_z|n> \label{aeqn}
\end{equation}
We have dropped all terms involving two
derivatives of the slowly varying quantities such as $a_n(z)$, $|n>$
and the exponent. Using the eigenvalue equation, Eq. 
\ref{eq:eigenvalueeq}, we find
\begin{equation}
<m|\partial_z|n> = {<m|(\partial_z M)|n>\over \mu^2_n-\mu^2_m}
\ \ \ \ m\ne n
\end{equation}
The resulting solution for $a_m$ can be written as
\begin{equation}
a_m(z)= e^{-{1\over 2}\int_0^z dz' {(\partial_{z'}\omega_m)/
\omega_m}}b_m(z) \label{wkbpre}
\end{equation}
where $b_m(z)$ satisfies
\begin{equation}
\partial_z b_m = {1\over \omega_m}\sum_{n,n\ne m} b_n\omega_n
{<m|(\partial_z M)|n>\over \mu^2_m-\mu^2_n}\sqrt{{\omega_n(0)\omega_m(z)
\over \omega_n(z)\omega_m(0)}}e^{i\int_0^zdz'(\omega_n-\omega_m)} 
\label{beqn}
\end{equation}

We pause to assess the result if we simply assumed $b_m(z)$ were 
constant. The pre-factor in Eq. \ref{wkbpre} is integrated via \ba 
exp\left( -{1\over 2}\int_0^z dz' {(\partial_{z'}\omega_m)/
\omega_m} \,\right)  = \sqrt{  \frac{ \omega_{m}(0 )}{\omega_{m}(z)   } }, 
\ea which is the well-known WKB $1/\sqrt{k(z)}$ factor cited earlier.  
In effect we have derived this factor and pushed the remaining 
derivatives in the equation for $b_{m}(z)$, Eq. \ref{beqn}.
It is clear that this exponential term is approximately equal to one.

We note that the equation for $b_{m}$ is very similar for the equation 
\ref{aeqn} for $a_{n}$. This sets up an iterative scheme.  Since slowly 
varying $\omega_n \sim \omega$, we can approximate Eq. \ref{beqn} as
\begin{equation}
\partial_z b_m \approx \sum_{n,n\ne m} b_n
{<m|(\partial_z M)|n>\over \mu^2_m-\mu^2_n}
e^{i\int_0^zdz'(\omega_n-\omega_m)}\ , 
\label{eq:bm}
\end{equation} plus small corrections in a series expansion of 
$|\omega-\omega_n|/\omega<<1$. 
For the cosmological applications, we consider here, the parameters 
ranges are such that the correction terms are extremely small.

The value of $b_m(L)-b_m(0)$ gives
an estimate of the transition amplitude between the two local 
eigenstates after the wave has propagated a distance $L$. Let us now 
estimate this with dimensional arguments. In the adiabatic limit, $b_n$,
as well as the term  
$ {<m|(\partial_z M)|n>/ (\mu^2_m-\mu^2_n)}$, 
varies slowly compared to the exponential. Hence we may take these to 
be approximately constant along the path. We can now integrate Eq. \ref{eq:bm} 
to obtain the transition amplitude. We find that $b_m(L)-b_m(0)$ is of order 
$\omega/(L\Delta\mu^2)$ as long as  
we ignore the case of resonant mixing, where the
two eigenmodes come very close to one another at some point along the path.
Hence we find that 
in the limit $L>> \omega/\Delta\mu^2$ the transition probability is 
negligible: this is the usual justification for the adiabatic 
approximation when the internal dynamical time scale is very short 
compared to the time scale for changes of parameters.


\subsubsection{Estimating the Transition Probability for Resonant Mixing}

We turn to estimating  the transition probability in the case of  
resonant mixing. We assume that 
the plasma frequency changes linearly along the  
path
according to the relation Eq. \ref{eq:linear}. We again take
the direction of the background magnetic field fixed so that we  
consider the two state mixing problem.
In this case we find, for example,
\begin{equation}
\partial_z b_2 = b_1{\alpha\over 2}{2g_{\phi} {\cal B}\omega\over  
\left[(C-\alpha z)^2
+ (2g_{\phi} {\cal B}\omega)^2\right]}e^{-i\int_0^z{dz'\over  
2\omega}(\mu^2_+-\mu^2_-)}
\label{eq:b2p}
\end{equation}
Now if we assume that the exponential is the most rapidly varying factor
on the right hand side of the above equation then we can replace the
coefficient by its average value along the path. We then find,
\begin{eqnarray}
b_2(L) &\approx& b_2(0) + i b_1(0)
{\omega\over L}\left[\tan^{-1}{m_\phi^2 - \omega_p^2(0)\over 2g_{\phi}  
{\cal B}\omega} -
\tan^{-1}{ m_\phi^2 - \omega_p^2(L) \over
2g_{\phi} {\cal B}\omega}\right]\nonumber\\ &\times &
\Bigg({e^{-i\int_0^Ldz'(\mu^2_+-\mu^2_-)/2\omega}
\over \sqrt{(m_\phi^2-\omega_p^2(L))^2+ (2g_{\phi} {\cal B}\omega)^2  
}}\nonumber\\ &-&
{1\over \sqrt{(m_\phi^2-\omega_p^2(0))^2+ (2g_{\phi} {\cal B}\omega)^2}  
}\Bigg)
\label{eq:transition}
\end{eqnarray}
where we have used
$$\left[{2g_{\phi} {\cal B}\omega\over (m_\phi^2-\omega_p^2(z))^2+  
(2g_{\phi} {\cal B}\omega)^2}\right]_{\rm av}
= {1\over L\alpha}\left[\tan^{-1}{m_\phi^2 - \omega_p^2(0)  \over  
2g_{\phi} {\cal B}\omega} -
\tan^{-1}{  m_\phi^2 - \omega_p^2(L)\over
2g_{\phi} {\cal B}\omega}\right]
$$
It is clear that the second term on the right hand side of Eq.
\ref{eq:transition} is small
and hence the transition probability from one state to another is
negligible as long as $L$ is sufficiently large such that
\begin{equation}
L>> {\omega\over |\mu^2_+-\mu^2_-| }\ .
\label{eq:condition_ad}
\end{equation}
However this condition is not general and is applicable only in the  
limit
of large $\omega$. The basic problem is that close to the region
where $\omega_p=m_\phi$, our original assumption that the coefficient of
the exponential in the right hand side of
Eq. \ref{eq:b2p} varies slowly compared to the exponential term
is not correct. We see this by comparing the magnitudes of the  
logarithmic
derivatives of these two factors. We find
$$D_1 = {d\over dz} \log \left[(C-\alpha z)^2 + (2g_{\phi} {\cal B}\omega)^2   
\right]
= {-2\alpha (C-\alpha z)\over (C-\alpha z)^2 + (2g_{\phi} {\cal  
B}\omega)^2}
$$
$$D_2 = {d\over dz}\log e^{-i\int_0^z{dz'\over 2\omega}(\mu^2_+-\mu^2_-)}
= {1\over 2\omega} \sqrt{(C-\alpha z)^2 + (2g_{\phi} {\cal B}\omega)^2}
$$
The assumption, $|D_1|<< |D_2|$,
implies
\begin{equation}
{|2\alpha (C-\alpha z)|\over (C-\alpha z)^2 + (2g_{\phi} {\cal  
B}\omega)^2}
<<{1\over 2\omega} \sqrt{(C-\alpha z)^2 + (2g_{\phi} {\cal B}\omega)^2}
\label{eq:condition1}
\end{equation}
In regions where $|C-\alpha z|>> 2g_{\phi} {\cal B}\omega$ this  
condition
demands
$$ |\alpha| << {1\over 4\omega}(C-\alpha z)^2\ .$$ Since $(C-\alpha z)$
is of order $m_\phi^2$ this condition implies $$ |\alpha| << {1\over  
4\omega}
m_\phi^2\ .$$ We next consider the limit where $|C-\alpha z|$ is small
or comparable to $2g_\phi {\cal B}\omega$.
We test the condition, Eq. \ref{eq:condition1}, 
at the point $|C-\alpha z|= \beta (2g_{\phi}  
{\cal B}\omega)$.
In this case we find
$$|\alpha| << {(\beta^2+1)^{3/2}\over \beta} (g_{\phi} {\cal  B})^2\omega$$
This condition is most stringent for very small values of $\beta$.  
However in this region, as we show below, the transition probability between
instantaneous eigenstates continues to be very small provided the  
condition
Eq. \ref{eq:condition_ad} is obeyed. If we take $\beta$ of order unity
then this condition implies
$|\alpha|<<(g_{\phi} {\cal B})^2\omega$, which can be expressed as,
\begin{equation}
L >> {|\omega_p^2(L) - \omega_p^2(0)|\over (g_{\phi} {\cal  
B})^2\omega}\ ,
\label{eq:condition_ad1}
\end{equation}
which is
rather stringent for small frequencies and magnetic fields. In regions
where this condition is violated we can no longer ignore the transition
between different instantaneous eigenstates.

We can obtain an estimate of the transition probability in
regions where Eq. \ref{eq:condition1} is violated.  Assume that
in these regions the exponential term in Eq. \ref{eq:b2p} varies much
slowly compared to the coeffient. We then replace the exponential
with its average over the path. We perform the integration over the
region where $C-\alpha z$ ranges from $\beta (2g_{\phi} {\cal  
B}\omega)$ to
$-\beta (2g_{\phi} {\cal B}\omega)$. In this region we find
\begin{equation}
\Delta b_2 = \tan^{-1}\beta \left[b_2
e^{-i\int_0^z{dz'\over 2\omega}(\mu^2_+-\mu^2_-)}
\right]_{\rm av}
\end{equation}
The exponential term gives a contribution of order unity. It follows  
that
the transition probability between different eigenstates is  
non-negligible, unless $\beta<<1$. 

\paragraph{Summary}
The above results can summarized by stating that in the case of
resonant mixing the adiabatic
approximation is valid only if both the equations
Eq. \ref{eq:condition_ad} and Eq. \ref{eq:condition_ad1} are
satisfied. If either of these equation is not respected then the
transition probability between instantaneous eigenstates cannot
be ignored.

\section{Applications} The basic aim of the present paper is to make
detailed predictions for polarization observables due to pseudoscalar
photon mixing, which can be tested in future astrophysical and
cosmological observations. Tests may involve CMBR, polarization
observations from distant sources, or propagation of 
electromagnetic waves through regions of strong magnetic fields such
as the pulsar magnetosphere or Active Galactic Nuclei. 
The resonance phenomenon may play an
important role in many of these situations since the waves propagate
through regions with large variations in plasma density. A detailed study
of all this phenomenon is beyond the scope of the present paper. Here
we shall confine ourselves to simple estimates. 

We first determine the
range of frequencies for which the adiabaticity condition is satisfied
in the case of resonance in the supercluster
magnetic fields. Using equation \ref{eq:condition_ad1}, with $B=0.1\ \mu$G
\cite{vallee},
$n_e=10^{-6}$ cm$^{-3}$, $L=10$ Mpc and the current limit on the coupling 
$g_\phi=6\times 10^{-11}$ GeV$^{-1}$, we find that adiabaticity is satisfied
if $\omega>> 5\times 10^{-3}$ eV. 
Hence we expect a significant effect at optical
frequencies but negligible at radio. We should point out that if 
resonance occurs it gives a considerably enhancement of the mixing 
phenomenon in comparison to the case of uniform background. For the
latter the mixing probability is limited by the oscillation
length $l$ in the medium and is proportional to $(gBl)^2$ \cite{CG94}.
For low frequencies this is very small and gives a significant effect only
if $\omega>5$ eV.  On the other hand for a spatially
varying medium if $\omega_p=m_\phi$ somewhere along the path, the mixing
probability is of order unity as long as the adiabaticity constraint 
Eq. \ref{eq:condition_ad1}
is satisfied. This constraint is satisfied at much smaller frequencies
in comparison to
what is required to obtain a significant effect if $\omega_p\ne m_\phi$ 
all along the path. 

For the galactic magnetic fields $B\approx 3\ \mu G$ and plasma density
$n_e\approx 0.03$ cm$^{-3}$ and distance scale 50 Kpc, we find that 
adiabaticity condition is satisfied only for $\omega >> 50 $ eV. Hence here
it is satisfied only for ultraviolet frequencies. 
Hence we can expect significant effects due to 
pseudoscalar-photon mixing at such high frequencies. 

The effects of resonance are most dramatic if  
$(gBl)^2<<1$, where $l$ is the oscillation length evaluated at some point
far away from resonant region. In this case the resonant effects dominate.
For supercluster magnetic fields this is found to be the case for a wide range
of frequencies $5\times 10^{-3} < \omega < 1$ eV (here the lower limit
is determined by the adiabaticity condition). In this case   
the results Eq. \ref{eq:reso_coherency} apply and the wave becomes almost
entirely linearly polarized after crossing the resonant region, 
independent of its state of polarization at origin. If the direction
of the magnetic field remains fixed during propagation then its state 
of polarization is well described by Eq. \ref{eq:reso_coherency}. In
general, however, the magnetic field will also twist along the path. In 
this case the wave will also acquire circular polarization during 
propagation. This phenomenon is discussed in a separate paper
\cite{Das_etal04}. If the frequency is large, such that $(gBl)^2>>1$,
then the effect of pseudoscalar-photon mixing is large irrespective 
of whether resonance occurs or not. 

The pseudoscalar-photon mixing has been proposed \cite{PJ} 
as a possible explanation
of the observed large scale alignment of optical polarizations from 
distant quasars \cite{Hutsemekers,Sarala}. The possibility of resonance
in supercluster magnetic fields further enhances the parameter space
over which this explanation may be applicable. This explanation can be further 
tested by observing the spectral dependence of all the Stokes parameters.
Pseudoscalar-photon mixing also predicts correlations between these
parameters, which can be tested in future observations. 

The resonance phenomenon may also considerably enhance
the production of pseudoscalars due to their mixing with photons as the
electromagnetic wave propagates through magnetics fields in astrophysical
objects such as the active galactic nuclei, pulsars, magnetars, 
galactic clusters etc. If we assume resonant production of pseudoscalars
from active galactic nuclei at optical frequencies then these pseudoscalars
can convert back into photons during propagation in the local supercluster.
It is reasonable to assume that the direct photon flux, integrated
over the entire source, would not be strongly polarized. Hence the observed
polarization may be determined dominantly due to the conversion of 
pseudoscalars into photons in the local magnetic field. This may 
explain the observed alignment of optical polarizations
from large redshift $z>1$ quasars over the entire sky \cite{Sarala} 
as well as their correlation with the supercluster equatorial plane 
\cite{Hutsemekers}.

\subsection{Cosmic Microwave Background Radiation} 
We next consider the effect of supercluster magnetic field on the CMBR due
to its mixing with pseudoscalars. Here even a small effect may be observable
due to the high precision with which CMBR has already been measured. 

{\it Intensity Effects:}
We first consider the intensity 
on the CMBR due to decay into pseudoscalars. Let $P(\theta,\phi,E)$
be the probability of decay  
into pseudoscalars, where $E$ is the energy of the photon 
and $(\theta,\phi)$ are
the angular coordinates. Let the original spectrum be denoted by $f(E;T)$
and the distorted spectrum due to mixing with pseudoscalars 
by $f'(E,T')$. These 
are related by 
\be f'(E;T') = f(E;T)(1-P(\theta,\phi,E)) \ee
Assuming that $P(\theta,\phi,E)<<1$ we find
\be T'(\theta,\phi,E) \approx T(\theta,\phi) - {P(\theta,\phi,E) T_0^2\over
E}(1-\exp(-E/T_0)) \ee
where $T_0$ is the mean CMBR temperature. As expected mixing with pseudoscalars
will produce both a frequency and angular dependence of the temperature. 
To be consistent with observations we expect $P(\theta,\phi,E)$ 
can have a maximum value of order $10^{-5}$. 

 {\it Angular Distributions:} 
It is interesting to speculate that the mixing
with pseudoscalars in the local supercluster might give rise to prefered 
orientation of the CMBR quadrupole
and octupole \cite{Oliveira,Eriksen} due to a possible angular
dependence of the mixing probability. This may also explain
why higher multipoles are aligned with the CMBR dipole \cite{JainRalston2004}.
``Cosmic variance", invoked to explain away magnitude puzzles, cannot
credibly explain the multiple coincidence of multipole directions.  
Hence an explanation in terms of local foreground effects is attractive.  
For consistency the mixing 
probability has to be of order $10^{-5}$, which is roughly the strength
of these multipole moments. Current limits on the pseudoscalar 
photon coupling do not rule out this possibility. 
However if local effects are the source, why should they be of the same order, 
multipole by multipole, as cosmological CMBR fluctuations?  
Interaction with a coherent pseudoscalar field might be the explanation
with the fewest number of arbitrary assumptions, although some 
fundamental revision of assumptions in cosmology might be needed to 
reconcile everything observed.
Nevertheless it is interesting to investigate pseudoscalar-photon mixing
in supercluster magnetic fields further
since it may provide more stringent limits on the coupling $g_\phi$.
Pseudoscalar-photon mixing will also give rise to a spectral dependence to the
CMBR temperature. Hence the effect has to be sufficiently small so that
it does not conflict with the observed agreement \cite{Firas}
of CMBR with black body radiation formula.

{\it Polarization:}
Pseudoscalar-photon mixing can also generate polarization of the CMBR radiation.
If the mixing probability is of order $10^{-5}$, which is required
to explain the alignment of quadrupole and octupole, then it will generate
polarization of the same order of magnitude. Hence if the 
polarization is observed to be much smaller than this, it might be possible
to impose further limits on the pseudoscalar photon
coupling. The pseudoscalar-photon mixing can also change the  
CMBR polarization parameters without changing the overall 
degree of polarization.
This effect is in general much larger. However if the degree of polarization
of CMBR is small, as expected, any change in the polarization will  
be observable by the current detectors only if it is of order unity. 
Hence here we require a relatively large contribution in order for new effects
to be observable. The shift in polarization is of the order of the phase 
acquired by the wave after propagating a distance $L$ \cite{PJ}.
For uniform medium this is of order $(g_\phi B)^2lL$, 
where $l$ is the oscillation
length. For the Virgo supercluster parameters and microwave frequencies
of order 100 GHz we find that the phase is of order $0.1$
for the coupling $g_\phi=6\times 10^{-11}$ GeV$^{-1}$. 
This phase will generate all the Stokes parameters with relative
strength of order $0.1$, assuming the current limit on the 
coupling $g_\phi$. Hence this contribution is small but 
it may be observable in future detectors. 
Similar results
are found if we take into account the fluctuations in the plasma
density \cite{PJ} or propagation through intergalactic medium \cite{Csaki}.

 {\it Condensate:}
A pseudoscalar condensate can
also affect the CMB polarization \cite{HS92}. In this case the rotation
of polarization is equal to $g_\phi\Delta\phi$, where $\Delta\phi$ is
the total change of the pseudoscalar field along the trajectory of the
electromagnetic wave. This effect, however, cannot generate circular
polarization. This effect is independent of frequency and will also
affect radio wave polarizations from distant sources \cite{PJ,JainRalston2004}. 
If this pseudoscalar field distribution is anisotropic,
it will lead to an anisotropy in both the CMB and radio 
polarizations \cite{Jain99}. Future observations can provide stringent
limits on this effect.

\section{Dark Energy} 
  {\it Dark energy} has come to denote a cosmic energy-momentum tensor
  of the vacuum. The source of dark energy is unknown.  Unfortunately,
  the traditionally minimal option to employ one universal cosmological
  constant appears less and less credible.
  It appears more likely that dark energy is the
  gravitational trace of a {\it field}, somehow chaperoning the evolution
  of Big Bang pressures, densities and phase transitions predicted by
  particle physics \cite{Ratra}. 
  The need for a causal inflaton field is rather
  clearly manifested in the high uniformity of the cosmic microwave
  background (CMB) radiation.

  By now, cosmology as a whole perhaps cannot do without a dark
  energy-associated field.  Yet of all interactions, exploring the
  Universe with gravity has a problem, in that gravity is the finest
  example where the coupling to fields is unknown!  Unconventional as the
  remark may seem, in quantum theory there exists no way to find
  independently that part of the energy momentum tensor which couples to
  gravity.  The rules of coupling are unknown, and if at all knowable,
  appear to hinge on ultra-high energy physics not experimentally
  testable.  The method of using gravity as a probe, when it is not
  really known what gravity couples to, flies in the face of the long
  successful tradition of using perturbatively stable interactions to
  study new situations.   It seems ironic that entire fields are built
  using gravity to explore the Universe's evolution when gravity is the
  least understood interaction.

If we assume that dark energy is associated with a scalar field $\phi$, 
such a field will
also have a coupling to electromagnetism given in Eq. \ref{Ldefined}.
Hence this field might 
produce all the physical effects we discuss in this paper.
Couplings of dark energy to ordinary matter fields as well as their
limits are discussed by Carroll \cite{Carroll98}. 
Yet the mass of such a conventional dark energy 
field is expected to be of order $10^{-33}$ eV, and then much smaller
than the intergalactic plasma frequency. Assuming such values we
do not expect the conditions for resonance to be applicable. One
may, however, consider generalized models of dark energy. Indeed 
it turns out that by invoking a false vacuum it is possible to explain
both dark matter and dark energy in terms of the invisible axion
\cite{Barr,Jain05}. In this case the mass of the background 
field can be much larger than currently assumed.
Hence we cannot rule out the possibility that conditions
for resonance may be applicable in some cases.

\section{Summary}

If one were to approach light and pseudoscalar field mixing  
``lightly,'' then there is a sequence of facile, dimensionally based  
arguments that could be used. First, since $g_{\phi}$ has dimensions of  
inverse mass, and electrodynamics is a theory with no scale, one might  
claim that all effects were relatively proportional to  
$g_{\phi}\omega$, and must vanish for $g_{\phi}\omega<<1$.  This is  
false: the effects are cumulative, and observable for exceedingly small  
$g_{\phi}<<10^{-12}$ GeV$^{-1}$.   Next one might observe that with a  
background field $\vec {\cal B}$, the effects must be of relative size  
$g_{\phi}{\cal B}/\omega$, and vanish for large $\omega$.  This is  
again false, as there are several other scales, sometimes $\omega$  
occurs in the numerator, and we find that for realistic parameters  
there are observable effects persisting all the way from radio to  
optical frequencies.

Then we return to our original goal of probing pseudoscalar field with light.   
It is certainly very interesting that light can be dimmed by  
interactions with pseudoscalars.  Yet in comparison the variety and  
variability of polarization-based observables seems almost unlimited.   
We believe that as polarization observations accumulate, more and more  
anomalies will appear.  It would be gratifying to have data so that a  
new and systematic study of pseudoscalars via a coupling that is stable  
under perturbation theory can commence.

\bf {Acknowledgments:}
Work supported in part under
Department of Energy grant number DE-FG02-04ER41308.

~

~

\end{document}